\documentclass[review, 11pt,sort&compress]{elsarticle}
\usepackage[utf8]{inputenc}
\usepackage[toc,page]{appendix}
\usepackage{amssymb}            
\usepackage{babel}
\usepackage{graphicx}
\usepackage{booktabs}
\usepackage{xcolor}
\usepackage{gensymb}
\usepackage{amsmath,amsfonts,amsthm,bm} 
\usepackage{mathrsfs} 
\usepackage{geometry}
\usepackage{gensymb}
\usepackage{multirow}
\usepackage[skip=6pt plus1pt, indent=40pt]{parskip}
\usepackage{setspace}
\usepackage{chngcntr}
\usepackage{upgreek}

\journal{Acta Biomaterialia}
\usepackage{graphicx}
\usepackage{caption}
\usepackage{subcaption}
\usepackage{float}
\usepackage{hyperref}
\hypersetup{colorlinks=true, linkcolor=blue, citecolor=black, filecolor=magenta, urlcolor=black}
\usepackage{soul}
\usepackage{accents}
\usepackage{color,soul}
\usepackage{color}
\usepackage{xcolor,soul}

\begin{document}
\begin{frontmatter}

\title{Impact of pH and chloride content on the biodegradation of magnesium alloys for medical implants: An in vitro and phase-field study}

\author[1] {Sasa Kovacevic\corref{cor1}}
\ead{sasa.kovacevic@eng.ox.ac.uk}

\author[2] {Wahaaj Ali}

\author[1]{Tushar Kanti Mandal} 

\author[1] {Emilio Mart\'inez-Pa\~neda\corref{cor1}}
\ead{emilio.martinez-paneda@eng.ox.ac.uk}

\author[2,3] {Javier LLorca\corref{cor1}}
\ead{javier.llorca@upm.es,javier.llorca@imdea.org}

\cortext[cor1]{Corresponding authors}

\address[1]{Department of Engineering Science, University of Oxford, Oxford OX1 3PJ, UK}
\address[2]{IMDEA Materials Institute, C/Eric Kandel 2, 28906 Getafe, Madrid, Spain}
%\address[3]{Department of Material Science and Engineering, Universidad Carlos III de Madrid, 28911 Leganes, Madrid, Spain}
\address[3]{Department of Materials Science, Polytechnic University of Madrid, E.T.S. de Ingenieros de Caminos, 28040 Madrid, Spain}
 
\begin{abstract} \vspace{-0.7mm}
% Max 250 words.
\noindent The individual contributions of pH and chloride concentration to the corrosion kinetics of bioabsorbable magnesium (Mg) alloys remain unresolved despite their significant roles as driving factors in Mg corrosion. This study demonstrates and quantifies hitherto unknown separate effects of pH and chloride content on the corrosion of Mg alloys pertinent to biomedical implant applications. The experimental setup designed for this purpose enables the quantification of the dependence of corrosion on pH and chloride concentration. The \textit{in vitro} tests conclusively demonstrate that variations in chloride concentration, relevant to biomedical applications, have a negligible effect on corrosion kinetics. The findings identify pH as a critical factor in the corrosion of bioabsorbable Mg alloys. A variationally consistent phase-field model is developed for assessing the degradation of Mg alloys in biological fluids. The model accurately predicts the corrosion performance of Mg alloys observed during the experiments, including their dependence on pH and chloride concentration. The capability of the framework to account for mechano-chemical effects during corrosion is demonstrated in practical orthopedic applications considering bioabsorbable Mg alloy implants for bone fracture fixation and porous scaffolds for bone tissue engineering. The strategy has the potential to assess the \textit{in vitro} and \textit{in vivo} service life of bioabsorbable Mg-based biomedical devices.
\end{abstract}
\end{frontmatter}
\vspace{-5mm}
\begin{spacing}{1.2}
\section*{Statement of significance} \label{sec} \vspace*{-1.7mm}
% less than 120 words
\noindent This investigation quantifies hitherto unknown individual effects of pH and chloride concentration on the corrosion kinetics of Mg alloys for biomedical applications. A dedicated experimental study is designed to isolate the separate contributions of pH and chloride concentration to the corrosion performance of these alloys. The findings reveal that pH is a dominant factor and plays a major role in the corrosion of bioabsorbable Mg alloys. Slight variations in chloride concentration do not affect the corrosion of bioabsorbable Mg alloys. The dependence of corrosion on pH, chloride concentration, and mechanical fields is integrated into a phase-field model applicable to both buffer-free and buffer-regulated media. The methodology provides a mechanistic tool for predicting the lifespan of bioabsorbable Mg-based biomedical devices.
\end{spacing}
\section{Introduction} \label{sec1}

Bioabsorbable Mg alloys have great potential for biomedical devices that gradually dissolve \textit{in vivo} with time. They neither hinder natural tissue growth nor require a second surgical intervention for device removal \cite{Li2020, BAIRAGI2022}. Ideally, the degradation of these devices should align with the rate of bone/tissue growth. Currently, bioabsorbable Mg-based implants have received commercial approval for orthopedic and cardiovascular applications, while others are undergoing clinical trials \cite{ZHAO2017, HAN2019}. Despite these successful applications, the practical use of Mg-based biomedical devices remains limited. A key impediment to the proliferation of Mg-based implants is the propensity for Mg to rapidly degrade in body fluids and exhibit susceptibility to mechanical loading in the case of load-bearing devices \cite{KIRKLAND2012, Jafari2015}. This fast degradation can compromise the mechanical integrity of implants to fulfill their designed requirements during the healing process, potentially leading to premature implant failures that can have life-threatening consequences \cite{Eliaz2018}. The quantification of the key factors that contribute to the corrosion of bioabsorbable Mg alloys in biological fluids and the development of simulation tools to assess the evolution of corrosion over time and the corresponding degradation in mechanical properties would help prevent such scenarios.

The performance of bioabsorbable Mg alloys is commonly assessed in \textit{in vitro} using various corrosive media prior to conducting \textit{in vivo} animal trials. These media include sodium chloride (NaCl) solutions, complex saline solutions, various types of simulated body fluids, cell culture media, and protein-contain solutions (cell culture medium with added proteins), among others \cite{MEI2020}. While the appearance of various inorganic ions, organic compounds, and proteins can inhibit or accelerate Mg corrosion \cite{XIN2008_agressiveIons, ZENG2014, MEI2019}, it is generally deemed that the amount of chloride content and the pH have a particularly strong effect \cite{Atrens2015_review}. Specifically, higher chloride ion (Cl$^-$) concentrations and lower pH values are ascribed to increased corrosion rates \cite{ZHAO2008, Wolf_Dieter2009}. This phenomenon originates from the well-known tendency of chloride ions to break down the corrosion product layer and the instability of this layer in solutions with low pH. The amount of Cl$^-$ concentrations in the above-mentioned \textit{in vitro} corrosive environments ranges from 0.085 M to 0.154 M \cite{MEI2020}. Besides the difference in chloride content, these media possess various pH levels controlled with different amounts of buffering agents. For instance, an unbuffered NaCl solution tends to have a pH of around 10.5, whereas buffered NaCl and other more complex corrosive media commonly have pH values ranging from 5.5 to 8.5 \cite{Johnston_Atrens2015, LAMAKA2017}. The extensive use of these corrosive environments to quantify the corrosion resistance of bioabsorbable Mg alloys shows significant inconsistencies in reported corrosion rates \cite{MUELLER2010, Xin_2010}. These discrepancies are typically attributed to differences in the composition of the test media \cite{XIN2011_review, Johnston_Atrens2018_review}. It remains unclear whether these discrepancies arise from variations in chloride concentration or pH values. The synergistic effect of chloride concentration and pH has been previously investigated in solutions with either one of these being fixed \cite{SONG1997_pH, XIN2008_agressiveIons, WANG2010_chloride, Taltavull2013, Johnston_Atrens2015, LAMAKA2017, CUI2017, MEI2019} or concentrations outside the typical range for biomedical applications \cite{ZHAO2008, Wolf_Dieter2009}. Therefore, the separate contributions of pH and chloride concentration to the corrosion kinetics of Mg alloys for biomedical applications remain an unsolved dilemma despite their significant roles as driving factors in Mg corrosion. Moreover, the existing numerical models \cite{GROGAN2014, SHEN2019, ZELLERPLUMHOFF2021, Bajger2017, GARTZKE2020, BARZEGARI2021, HERMANN2022, KOVACEVIC2023, XIE2024, ZHANG2024, Hermann2025} do not account for the dependence of corrosion on pH and chloride concentration, and as such, they are not capable of assessing the corrosion performance of Mg alloys in corrosive media with variable chloride and pH values.

This work demonstrates and quantifies the effect of pH and chloride concentration on the corrosion of bioabsorbable Mg alloys pertinent to biomedical implant applications. A dedicated experimental study is tailored to isolate the individual contributions of pH and chloride content to corrosion kinetics. The simplest corrosive solutions in terms of NaCl are considered to avoid unwanted effects from other ions and compounds. The influence of pH on corrosion behavior is evaluated through \textit{in vitro} tests conducted in two different environments. One set of tests is performed in air without controlling the pH (buffer-free solution). Another set is carried out in an incubator supplied with a constant 5\% CO$_2$ atmosphere to maintain a stable pH value, an approach followed and recommended in the literature \cite{MEI2020, Atrens2015_review, Johnston_Atrens2015}. Chloride concentration is varied in each of these test environments. The amount of chlorine concentration encompasses values used in the aforementioned corrosive media for Mg corrosion tests for biomedical applications \cite{MEI2020}. Subsequently, a variationally consistent diffuse interface (phase-field) framework is developed applicable to both buffer-free and buffer-regulated solutions with sensitivities to pH, chloride content, and mechanical fields. Phase-field models have emerged as a promising approach for modeling corrosion damage at different length scales and in various materials and media \cite{KOVACEVIC2023, XIE2024, ZHANG2024, OKAJIMA2010, Liang2012, Mai2018, Chadwick2018, Ansari2018, Tsuyuki2018, LIN2021, Brewick2022, Cui2023, Makuch2024, martinez2024phase, KANDEKAR2024}. 

The outline of the paper is as follows. The experimental setup, test environments, and experimental methods are presented in Section \ref{sec2}. The phase-field model of biocorrosion with the dependence of corrosion on pH, chloride concentration, and mechanical fields is formulated in Section \ref{sec3}. The main kinematic variables necessary for model description are defined based on the underlying chemical reactions responsible for the corrosion of Mg alloys in the corrosive solutions considered. The motion of the metal$-$liquid interface follows the usual kinetic rate theory for chemical reactions. The role of mechanical fields in enhancing corrosion kinetics is assessed by means of Gutman's mechanochemical theory \cite{Gutman1988}. The model is calibrated against the \textit{in vitro} tests with varying pH and chlorine content in Section \ref{sec4}. After validation, the potential of this strategy to assess the degradation of bioabsorbable Mg implants subjected to mechanical loading is demonstrated in Section \ref{sec5} by considering two practical orthopedic applications: plates and screws for bone fracture fixation and porous bone scaffolds for bone repair. The findings from the \textit{in vitro} tests and numerical simulations, along with recommendations for future work, are discussed in Section \ref{sec6}. The manuscript ends with concluding remarks in Section \ref{sec7}.

\section{Materials and methods} \label{sec2}

\subsection{Materials and solutions} \label{sec21}

WE43MEO Mg alloy wires were used in the degradation experiments. The wires were manufactured at Meotec GmbH (Aachen, Germany) by indirect extrusion, followed by several cold drawing and annealing steps \cite{MGWIRE1}. The final diameter of the wires was 0.3 mm. Their nominal composition included 1.4-4.2\% Y, 2.5-3.5\% Nd, $<$ 1\% (Al, Fe, Cu, Ni, Mn, Zn, Zr) and balance Mg (in wt.\%). The mechanical properties of the wires were assessed by performing uniaxial tensile tests. The yield strength, the tensile strength, and the strain-to-failure of the wires were approximately 250 MPa, 320 MPa, and 9\% \cite{MGWIRE2}.

\begin{figure}[h!]
    \centering
    \includegraphics[width = 16 cm]{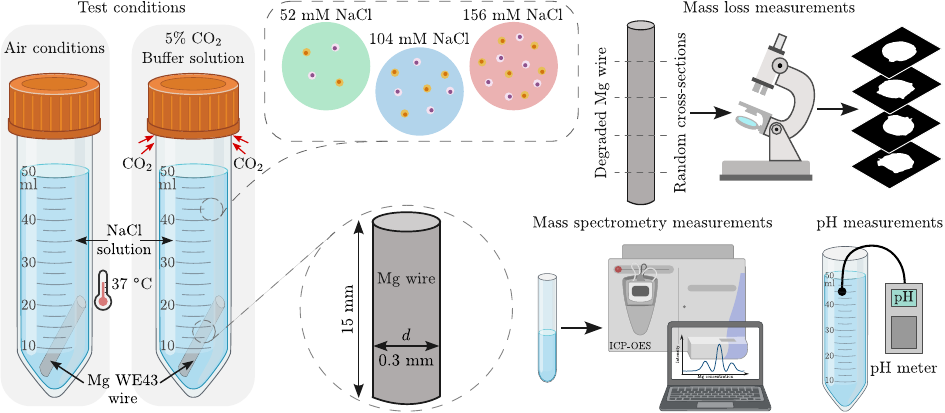}
     \captionsetup{labelfont = bf, justification = raggedright}
    \caption{Schematic illustration of the experimental setup for the Mg alloy wires immersed in different NaCl solutions.}
    \label{Fig1}
\end{figure}

The degradation tests were carried out on wires of 15 mm in length that were immersed in 50 mL of NaCl solution. The solution was prepared by dissolving NaCl in deionized water. The initial pH of the solution was 5.5. Two test environments were considered to evaluate the impact of pH: air conditions without controlling the pH level (buffer-free solution) and a bicarbonate buffer system containing 5\% CO$_2$ (buffer-regulated medium), Table \ref{table01}. The corresponding chemical reaction for Mg dissolution in NaCl-based solutions and bicarbonate buffer systems are provided in Eqs. (\ref{eqn1a})$-$(\ref{eqn2b}), Section \ref{sec31}. The effect of chlorine concentration on the degradation behavior of the wires was studied by varying the content of Cl$^-$ ions in the solution. Three NaCl solutions were prepared with different chloride concentrations for each test environment: 52 mM, 104 mM, and 156 mM NaCl. The solution containing 104 mM NaCl corresponds to human blood plasma and is a reference for comparison. The experimental setup is schematically depicted in Fig. \ref{Fig1}.

\begin{table}[t]
\centering
\begin{tabular}{ l l l l} 
 \hline
 
 Test environment & Initial pH & Final pH & Chloride content\\
\hline
Air conditions & \multirow{2}{4em}{5.5}  & \multirow{2}{4em}{10.5} & \multirow{2}{12em}{52 mM, 104 mM, 156 mM}\\
(buffer-free solution) & \\
5\% CO$_2$ conditions & \multirow{2}{4em}{5.5}  & \multirow{2}{4em}{6.2} & \multirow{2}{12em}{52 mM, 104 mM, 156 mM}\\
(buffer-controlled solution) & \\
\hline
\end{tabular}
\captionsetup{labelfont = bf,justification = centering}
\caption{Test conditions used to conduct the experiments.}
\label{table01}
\end{table}

\subsection{Experimental methods} \label{sec22}

The samples were inserted in a falcon tube filled with solution and incubated at 37 $\degree$C. In the air environment, the cap of the falcon tube was tightly sealed, Fig. \ref{Fig1}. This test condition increased pH from the initial value of 5.5 to a steady state value of $\sim$ 10.5, Eqs. (\ref{eqn1a})$-$(\ref{eqn1d}). In the 5\% CO$_2$ environment, the cap of the falcon tube was loosely closed, allowing CO$_2$ to diffuse into the solution. This buffering effect helped maintain the pH of the solution, Eq. (\ref{eqn2a}). Under these CO$_2$ conditions, the pH increased from the initial value of 5.5 to $\sim$ 6.2 and was kept at this level by a continuous supply of CO$_2$. The ratio of liquid volume to the wire surface area was $>$ 0.5 mL/mm$^2$, according to the ASTM G31-72 standard. During immersion time, experimental measurements were frequently taken in terms of mass loss, the average concentration of Mg ions in solution, and pH. Measurements were recorded after one, four, nine, and fourteen days of immersion in the air environment. In the 5\% CO$_2$ environment, readings were conducted at one, two, and three days of immersion. To ensure repeatability, three measurements were taken for each NaCl solution at each recording time.

Mass loss measurements were conducted to evaluate the degradation kinetics of the wires. After immersion, the samples were carefully removed from the NaCl solution and dried in air. They were then randomly cut into several pieces and fixed vertically in a plastic holder using tape. The holder was then embedded in an epoxy resin. The cut cross-sections were grounded and polished.  At least ten images of randomly selected cross-sections of the degraded Mg wires were studied using an optical microscope to ensure statistical significance. These images were processed with ImageJ software to determine the remaining area of the Mg wire, which is indicated in white in Fig. \ref{Fig1}. The average mass loss was obtained by subtracting the remaining uncorroded regions from the original areas (before degradation) for all randomly selected cross-sections. The morphology of the corrosion product layer was analyzed by a Back-Scattered Scanning Electron Microscope (Zeiss EVO MA15, Germany) at 20 kV, followed by the gold sputtering of polished cross-sections.

After removing the wires from the solution to compute mass loss, the remaining NaCl solution was mixed and additional measurements were carried out. The average concentration of magnesium ions (Mg$^{2+}$) in the solution was measured using an Inductively Coupled Plasma Optical Emission Spectrometer (ICP-OES iCAP PRO XPS, ThermoFisher). The dissolution of the corrosion product layer (Eqs. (\ref{eqn1c}) - (\ref{eqn1d})) was estimated by correlating the average concentration of Mg$^{2+}$ ions in the solution with mass loss measurements, as detailed below in Section \ref{sec4}. The bulk pH of the solution was measured using a pH meter (Hach Sension$^+$). The same procedures and steps were consistently followed for each test environment and the NaCl solutions considered. The experimental data of mass loss, the average concentration of Mg$^{2+}$ ions in solution, and bulk pH of the solution as a function of immersion time for the three NaCl solutions analyzed in both test environments are presented in Fig. \ref{Fig5}, along with model predictions. The experimental data and model predictions are discussed in Section \ref{sec4}. 

\subsection{Statistical analysis} \label{sec23}

The experimental measurements are expressed as mean value $\pm$ standard deviation, Fig. \ref{Fig5}. Microsoft Excel software is used for the statistical calculations.

\section{The phase-field model of biocorrosion} \label{sec3}

\subsection{Underlying electrochemistry} \label{sec31}

Magnesium dissolves in NaCl-based solutions via an electrochemical reaction with water \cite{ZHENG2014}
\begin{align}
\label{eqn1a}
& \text{Mg}_\mathrm{(s)} \rightarrow \text{Mg}_\mathrm{(aq)}^{2+} + 2e^- \text{ (anodic reaction)} \\
\label{eqn1b}
& \text{2H}_2\text{O}_\mathrm{(aq)} + 2e^- \rightarrow \text{H}_\mathrm{2(g)} \uparrow + \text{2OH}^-_\mathrm{(aq)}  \text{ (cathodic reaction)}\\
\label{eqn1c}
& \text{Mg}^{2+}_\mathrm{(aq)} + \text{2OH}^-_\mathrm{(aq)} \rightarrow \text{Mg(OH)}_\mathrm{2(s)} \downarrow   \text{ (layer formation)}\\
\label{eqn1d}
& \text{Mg(OH)}_\mathrm{2(s)} + \text{2Cl}^-_\mathrm{(aq)} \rightarrow \text{MgCl}_\mathrm{2(aq)} + \text{2OH}^-_\mathrm{(aq)} \text{ (layer dissolution)},
\end{align}
producing hydrogen gas and a porous corrosion product layer consisting of magnesium hydroxide ($\text{Mg(OH)}_{2}$) that forms on the exposed metal surface. The layer initiates at regions on the metal surface where local concentrations of Mg$^{2+}$ ions and hydroxide ions (OH$^{-}$) exceed the solubility product constant of Mg(OH)$_2$ \cite{Williams2014}. The formation rate of the Mg(OH)$_2$ layer considerably depends on the synergistic effect of pH and concentration of aggressive $\text{Cl}^-$ ions. The production of OH$^{-}$ ions in Eq. (\ref{eqn1b}) leads to an increase in pH that favors the formation and stabilization of the hydroxide layer, Eq. (\ref{eqn1c}). However, the presence of $\text{Cl}^-$ ions detains the formation of the corrosion product layer by increasing the solubility product constant of Mg(OH)$_2$ \cite{Williams2017}. The time required for this layer to form increases, leading to a faster corrosion rate. $\text{Cl}^-$ ions also react with the corrosion layer and form highly soluble magnesium chloride (MgCl$_2$), Eq. (\ref{eqn1d}). The hydroxide layer begins to dissolve and is prone to local breakdowns and thinning, creating layer-free areas that act as nuclei for localized (pitting) corrosion \cite{XIN2008}.

The bicarbonate buffer system maintains the pH by regulating the levels of carbonic acid ($\text{H}_2\text{CO}_3$), bicarbonate ions ($\text{HCO}_3^-$), and carbon dioxide (CO$_2$) via the following reaction
\begin{align}
\label{eqn2a}
& \text{CO}_{2} +\text{H}_{2}\text{O} \leftrightarrow \text{H}_{2}\text{CO}_3 \leftrightarrow \text{HCO}_3^- + \text{H}^+\\
\label{eqn2b}
& \text{Mg(OH)}_2 + 2\text{CO}_2 \rightarrow \text{Mg(HCO}_3)_2 \rightarrow \text{MgCO}_3 + \text{CO}_{2} +\text{H}_{2}\text{O}
\end{align}
The production of hydrogen ions (H$^+$) controls the pH value. The Mg(OH)$_2$ layer interacts with free CO$_2$ and produces Mg bicarbonate (Mg(HCO$_3$)$_2$), which is an unstable compound that quickly dissolves into Mg carbonate (MgCO$_3$).

\subsection{Kinematics and thermodynamics} \label{sec32}

\begin{figure}[h!]
    \centering
    \includegraphics[width = 16 cm]{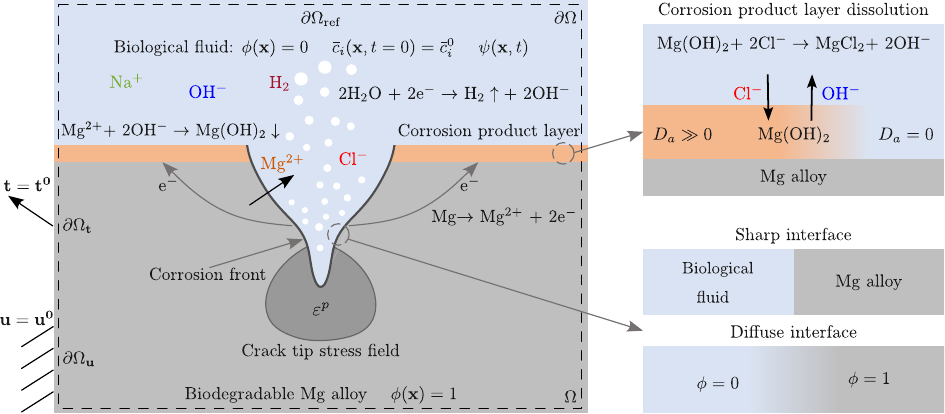}
    \captionsetup{labelfont = bf, justification = raggedright}
    \caption{Underlying electrochemistry and diffuse interface description of the liquid (biological fluid $\phi$ = 0) and solid (biodegradable Mg alloy $\phi = 1$) phases.}
    \label{Fig2}
\end{figure}

The system domain $\Omega$ comprises a biodegradable Mg alloy that acts as an electrode, an electrolyte that consists of positively and negatively charged aqueous ions, and a porous corrosion product layer that forms on the solid$-$liquid interface as the corrosion proceeds, Fig. \ref{Fig2}. The following concentrations describe the surrounding environment and the reactions in Eqs. (\ref{eqn1a})$-$(\ref{eqn2b}): the concentration of solidus species (Mg metal atoms) $c_{\text{Mg}}$ present solely in the metal (electrode) and the concentration of ionic species present in the electrolyte $c_{\text{Mg}^{2+}}$, $c_{\text{OH}^-}$, and $c_{\text{Cl}^-}$. Albeit sodium ions ($\text{Na}^+$) are present in a NaCl-containing solution, they are not included in the present model as they do not take part in Eqs. (\ref{eqn1a})$-$(\ref{eqn2b}). Yet, their effect on the motion of other ions in solution is implicitly incorporated through the electric conductivity of the electrolyte (Section \ref{sec334}). The concentration of $\text{MgCl}_2$ is not considered due to its high solubility in aqueous solutions \cite{Haynes2016}. The production of $\text{H}^+$ ions in Eq. (\ref{eqn2a}) is modeled as consumption of $\text{OH}^-$ ions. $\text{HCO}_3^-$ ions in Eq. (\ref{eqn2a}) and MgCO$_3$ in Eq. (\ref{eqn2b}) are not tracked in the present model as their effect on corrosion behavior is complex and yet to be understood \cite{XIN2008, XIN2011_review, Johnston_Atrens2018_review}. Given that the thin and highly porous hydroxide layer only affects the transport of ionic species in the narrow interfacial area, it is not explicitly included in the present model. Its effect on corrosion kinetics is integrated through an effective reaction rate that accounts for the consumption of Mg$^{2+}$ and OH$^-$ ions involved in forming the layer, Section \ref{sec346}. Although this simplification omits the impact of this layer on interfacial mass transport, it enables incorporating its influence on corrosion kinetics without the need to simulate it directly within the computational domain. The question of explicitly modeling the precipitation and dissolution of the corrosion product layer within the diffuse interface formulation has yet to be addressed. The sensitivity of the formation of the hydroxide layer on Cl$^-$ ions is implemented through the dependence of the solubility product constant of Mg(OH)$_2$ on the chloride concentration, Section \ref{sec347}.

A continuous phase-field variable $\phi(\mathbf{x},t)$ with a physical correspondence to the normalized Mg metal concentration $\phi = c_{\text{Mg}} V_\mathrm{m}$ is introduced to distinguish different phases and for describing the evolution of the corroding front: $\phi = 1$ represents the solid phase (Mg alloy constituted only of Mg metal atoms), $\phi = 0$ corresponds to the liquid phase (corrosive medium with no Mg metal concentration), and $0 < \phi < 1$ indicates the thin interfacial region between the phases (solid$-$liquid interface) where the electrochemical reaction takes place. $V_\mathrm{m}$ denotes the molar volume of the material. To simplify notation, the ionic species considered are described by the set of dimensionless concentrations as $\overrightarrow{\bar{c}} = \{\bar{c}_{1} = c_{\text{Mg}^{2+}} / c_0, \bar{c}_{2} = c_{\text{OH}^-} / c_0, \bar{c}_{3} = c_{\text{Cl}^-} / c_0\}$, where $c_0$ stands for the standard bulk concentration of electrolyte solution. The underlying electrochemistry and corrosion mechanisms considered are schematically given in Fig. \ref{Fig2}.

The free energy functional of the heterogeneous system in Fig. \ref{Fig2} includes the chemical $\mathcal{F}^\mathrm{chem}$, electric $\mathcal{F}^\mathrm{elec}$, gradient $\mathcal{F}^\mathrm{grad}$, and mechanical $\mathcal{F}^\mathrm{mech}$ free energy densities 
\begin{equation} \label{eqn3}
                   \mathscr{F} = \int_\Omega \Big[\mathcal{F}^\mathrm{chem}(\overrightarrow{\bar{c}}, \phi) + \mathcal{F}^\mathrm{elec}(\overrightarrow{\bar{c}},\psi) + \mathcal{F}^\mathrm{grad}(\nabla\phi) + \mathcal{F}^\mathrm{mech}(\phi, \nabla \mathbf{u}) \Big]\,\text{d}\Omega,
\end{equation}
where $\mathbf{u}(\mathbf{x},t)$ is the displacement vector to characterize the deformation of the material and $\psi (\mathbf{x},t)$ the solution potential that arises due to uneven distributions of charged particles. 

The total chemical free energy density $\mathcal{F}^\mathrm{chem}$ is decomposed into the chemical free energy density associated with the ionic species present in solution and the double-well free energy density associated with the phase-field variable \cite{Bazant2013}
\begin{equation} \label{eqn4}
\mathcal{F}^\mathrm{chem} (\overrightarrow{\bar{c}}, \phi) = c_0 R T \sum \bar{c}_i\ln \bar{c}_i + c_0\sum \bar{c}_{i}\mu_{i}^{\Theta} + \omega g(\phi),
\end{equation}
where $R$ is the universal gas constant, $T$ the absolute temperature (in Kelvin), $\mu_{i}^{\Theta}$ the reference molar chemical potential of ionic species $i$, and $g(\phi) = 16\phi^2(1-\phi)^2$ the double-well free energy function employed to describe the two equilibrium states for the solid ($\phi = 1$) and the liquid ($\phi = 0$) phases. $\omega$ in Eq. (\ref{eqn4}) is a constant that determines the energy barrier at $\phi = 1/2$ between the two minima at $\phi = 0$ and $\phi = 1$. The parameter $\omega$ is defined as $\omega =3\Gamma/(4\ell)$, where $\Gamma$ stands for the interfacial energy and $\ell$ the chosen nominal interface thickness \cite{KOVACEVIC2020}. Throughout this work, the summation sign denotes a summation over all ionic species present in the liquid phase, i.e., $\sum_{i = 1}^3$.

The electric $\mathcal{F}^\mathrm{elec}$ and interfacial $\mathcal{F}^\mathrm{grad}$ free energy densities are commonly given as
\begin{equation} \label{eqn5}
                   \mathcal{F}^\mathrm{elec}(\overrightarrow{\bar{c}},\psi) = c_0 F \psi \sum z_i\bar{c}_i \quad\mathrm{}\quad \mathcal{F}^\mathrm{grad}(\nabla\phi) = \frac{1}{2}\kappa|\nabla \phi|^2,
\end{equation}
where $F$ is Faraday's constant, $z_i$ the charge carried by component $i$, and $\kappa = 3\Gamma\ell/2$ the isotropic gradient energy coefficient \cite{KOVACEVIC2020}.

The mechanical free energy density $\mathcal{F}^\mathrm{mech}$ is only attributed to the solid phase ($\phi = 1$). Considering deformable elasto-plastic solids, the mechanical free energy density is additively decomposed into elastic $\mathcal{F}_e^\mathrm{mech}$ and plastic $\mathcal{F}_p^\mathrm{mech}$ components 
\begin{equation} \label{eqn8}
                   \mathcal{F}^\mathrm{mech}(\phi, \nabla \mathbf{u}) = h(\phi)(\mathcal{F}_e^\mathrm{mech} + \mathcal{F}_p^\mathrm{mech}),
\end{equation}
where $h(\phi) = \phi^3(6\phi^2 - 15\phi + 10)$ acts as a degradation function to account for the transition from the uncorroded Mg alloy ($\phi = 1$) to the completely corroded phase ($\phi = 0$). The elastic and plastic strain energy densities are defined as
\begin{equation} \label{eqn9}
                   \mathcal{F}_e^\mathrm{mech}(\nabla \mathbf{u}) = \frac{1}{2} \bm{\varepsilon}^{e} : \mathbf{C} : \bm{\varepsilon}^{e} \quad\mathrm{}\quad \mathcal{F}_p^\mathrm{mech} = \frac{\sigma_y \varepsilon_0}{N+1}\Big [ \Big(1 + \frac{\varepsilon^p}{\varepsilon_0} \Big)^{N+1} - 1 \Big ],
\end{equation}
where $\mathbf{C}$ is the rank-four elastic stiffness tensor and $\bm{\varepsilon}^{e} = \bm{\varepsilon} - \bm{\varepsilon}^{p}$ is the elastic strain tensor obtained by subtracting the plastic strain tensor $\bm{\varepsilon}^{p}$ from the total strain tensor $\bm{\varepsilon} = 1/2 (\nabla \mathbf{u} + (\nabla \mathbf{u})^T)$. The rank-four elastic stiffness tensor is described by the isotropic linear elasticity theory: $C_{ijkl} = \lambda \delta_{ij}\delta_{kl}+\mu(\delta_{ik}\delta_{jl}+\delta_{il}\delta_{jk})$, where $\lambda$ and $\mu$ are Lam\'e elastic constants. The plastic part of mechanical free energy density for the intact solid $\mathcal{F}_p^\mathrm{mech}$ in Eq. (\ref{eqn9}) is assumed to follow an isotropic power law hardening response with the von Mises theory of plasticity \cite{Simo1998}. Here, $\sigma_y$ is the yield strength, $\varepsilon_0$ the onset of strain for hardening, $N$ the hardening exponent, and $\varepsilon^p = \sqrt{2/3} \int_{0}^{t} | \dot{\bm{\varepsilon}}^{p} |\, dt$ the von Mises equivalent plastic strain. 

\subsection{Governing equations} \label{sec33}

\subsubsection{Solid-liquid interface evolution and electrode kinetics} \label{sec331}

The motion of the interface is governed by the electrochemical reaction at the metal$-$electrolyte interface. Following the convective Allen-Cahn equation \cite{ALLEN1979}, the evolution of the solid$-$liquid interface is written as  
\begin{equation} \label{eqn12}
\frac{\partial \phi}{\partial t} = - L\frac{\delta \mathscr{F}}{\delta \phi} - \mathbf{v} \cdot \nabla \phi  \quad \text{in}\quad\Omega; \quad\mathrm{} \kappa \mathbf{n} \cdot \nabla \phi = 0 \quad \text{on}\quad\partial\Omega.
\end{equation}
The first term is the diffusive transport of the interface associated with the interfacial free energy reduction. The second term is the convective transport of the interface due to the electrochemical reaction at the metal$-$electrolyte interface. The mobility parameter $L > 0$ needs to be large enough to retain a constant interfacial thickness but small enough so that the convective term is not overly damped. The dimensional analysis in Section \ref{sec34} provides the condition for its selection. The interface velocity due to the electrochemical reaction at the solid$-$liquid interface $\mathbf{v}$ is defined by following Faraday's second law 
\begin{equation} \label{eqn13}
\mathbf{v} = \frac{i}{c_\mathrm{s} F z} \mathbf{n_\phi},
\end{equation}
where $i$ is the current density, $z$ the number of electrons involved in the electrochemical reaction ($z = 2$), $c_\mathrm{s} = 1/V_\mathrm{m}$ the site density of the Mg alloy, and $\mathbf{n_\phi} = \nabla \phi/|\nabla \phi|$ the unit normal to the evolving solid$-$liquid interface pointing towards the solid phase. The current density generated by the electrochemical reaction is assumed to follow the usual kinetic rate theory for chemical reactions \cite{Jones1996}
\begin{equation} \label{eqn14}
\begin{aligned}
i = i_0\Big[\text{exp}\Big (\frac{\alpha_{\mathrm{a}} z F}{RT} (\eta + \Delta \eta_{\sigma}) \Big) - \text{exp}\Big(-\frac{(1-\alpha_{\mathrm{a}}) z F}{RT} \eta \Big)\Big] \quad\mathrm{}\quad \eta = \psi_\mathrm{s} - \psi - \Delta \psi^\mathrm{eq},
\end{aligned}
\end{equation}
where $i_0$ is the exchange current density, $\alpha_{\mathrm{a}}$ the anodic transfer coefficient, $\eta$ the total overpotential of the non-deformed electrode, and $\psi_\mathrm{s}$ the potential of the metal ($\psi_\mathrm{s} = 0$ V in this work). The equilibrium electrode-electrolyte potential difference $\Delta \psi^{\mathrm{eq}}$ and the increase in the overpotential due to the presence of mechanical fields $\Delta \eta_{\sigma}$ are given by the standard Nernst equation \cite{CIOBANU2007} and the theory of mechano-electrochemical interactions \cite{Gutman1988}
\begin{equation} \label{eqn15}
\begin{aligned}
\Delta \psi^\mathrm{eq} = E^{\Theta} + 2.303 \frac{RT}{F}\text{pH} - \frac{RT}{z F}\text{ln}\Big(\frac{c^\mathrm{s}_{\text{Mg}^{2+}}}{c^\mathrm{b}_{\text{Mg}^{2+}}}\Big) \quad\mathrm{}\quad \Delta \eta_{\sigma} = \frac{\sigma_\mathrm{h} V_\mathrm{m}}{z F} + \frac{RT}{z F}\text{ln}\Big (\frac{\varepsilon_p}{\varepsilon_0} + 1 \Big),
\end{aligned}
\end{equation}
where $E^{\Theta}$ is the standard electrode potential difference between reactants and products, $c^\mathrm{s}_{\text{Mg}^{2+}}$ the concentration of $\text{Mg}^{2+}$ ions at the electrode surface (metal$-$electrolyte interface), $c^\mathrm{b}_{\text{Mg}^{2+}}$ the concentration of $\text{Mg}^{2+}$ ions in the bulk solution, and $\sigma_\mathrm{h}$ the hydrostatic stress. The equilibrium potential difference $\Delta \psi^\mathrm{eq}$ depends on pH and the concentration of reaction products near the electrode surface. A more negative value of $\Delta \psi^\mathrm{eq}$ correlates with a higher corrosion rate. Mechanical fields affect only the anodic current density \cite{Gutman1988}. After substituting Eq. (\ref{eqn15}) into Eq. (\ref{eqn14}) and standard manipulation, the current density of the mechanically deformed electrode can be expressed as
\begin{equation} \label{eqn16}
\begin{aligned}
i = i_\mathrm{a} \text{exp}\Big( \frac{\alpha_\mathrm{a} \sigma_\mathrm{h} V_\mathrm{m}}{RT}\Big) \Big(\frac{\varepsilon_p}{\varepsilon_0} + 1\Big)^{\alpha_\mathrm{a}} - i_\mathrm{c},
\end{aligned}
\end{equation}
where $i_\mathrm{a} = i_0 \text{exp} (\alpha_{\mathrm{a}} z F \eta/(RT))$ and $i_\mathrm{c} = i_0 \text{exp}(-(1-\alpha_{\mathrm{a}}) z F \eta/(RT))$  are the anodic and cathodic current densities of the non-deformed electrode. Combining Eq. (\ref{eqn16}) with Eq. (\ref{eqn13}) returns the governing equation for the evolution of the solid$-$liquid interface
\begin{equation} \label{eqn17}
\frac{\partial \phi}{\partial t} = -L\Big(\frac{\partial \mathcal{F}^\mathrm{chem}}{\partial\phi} + \frac{\partial \mathcal{F}^\mathrm{mech}}{\partial\phi} - \kappa\nabla^2\phi \Big) - \frac{i}{c_\mathrm{s} F z}|\nabla \phi| \quad\mathrm{} \text{in }\Omega.
\end{equation} 
The interface motion is linearly proportional to the interfacial free energy reduction (diffusive transport) and nonlinearly with respect to the current density (convective transport). The latter exhibits nonlinear dependence on pH, solution potential $\psi$, and hydrostatic stress $\sigma_\mathrm{h}$. 

\subsubsection{Mass transport equation} \label{sec332}

The transport of ionic species in the electrolyte is subjected to the mass balance law
\begin{equation} \label{eqn18}
\frac{\partial \bar{c}_i}{\partial t} = - \nabla \cdot \mathbf{J}_i + R_i \quad \text{in}\quad\Omega; \quad\mathrm{}\quad \mathbf{J}_i = -M_i \nabla \Big(\frac{\delta \mathscr{F}}{\delta \bar{c}_i} \Big) = -D_i \Big(\nabla \bar{c}_i + \frac{z_i F}{R T} \bar{c}_i \nabla \psi \Big),
\end{equation}
where $\mathbf{J}_i$ is the electrochemical flux and $M_i$ the mobility parameter expressed using the Nernst-Einstein equation $M_i = D_i/(\partial^2 \mathcal{F}^\mathrm{chem} / \partial \bar{c}_i^2)$. Here, $D_i$ stands for the diffusion coefficient of ionic species $i$. On the boundary $\partial\Omega$: $\mathbf{n} \cdot \mathbf{J}_i = 0$. The total reaction rates $R_i = R_i^{\prime} + R_i^{\prime\prime}$ are decomposed into the local electrochemical rates $R_i^{\prime}$ due to the anodic and cathodic reactions at the solid$-$liquid interface (Eqs. (\ref{eqn1a}) and (\ref{eqn1b})) and the volumetric rates $R_i^{\prime\prime}$ associated with the precipitation and dissolution of the corrosion product layer, Eqs. (\ref{eqn1c}) and (\ref{eqn1d}). $R_i^{\prime\prime}$ also accounts for the production of H$^+$ ions and MgCO$_3$ in the case of the buffer solution, Eqs. (\ref{eqn2a}) and (\ref{eqn2b}). The accompanying reaction rates $R_i^{\prime}$ and $R_i^{\prime\prime}$ are defined below in Section \ref{sec346}.

\subsubsection{Electric potential distribution} \label{sec334}

By assuming that the solution is electrically neutral ($\sum z_i \bar{c}_i = \sum z_i \partial \bar{c}_i/\partial t = 0$), the charge conservation law reduces to the current balance equation
\begin{equation} \label{eqn22}
    \nabla \cdot c_0 F \sum \mathbf{J}_i z_i = c_0 F \sum R_i z_i + Q,
\end{equation}
where $c_0 F \sum \mathbf{J}_i z_i$ is the net current density defined as the sum of electrochemical flux of all ionic species, $c_0 F \sum R_i z_i$ the charge that enters or leaves the system due to the chemical reactions, and $Q = c_0 F \mathbf{v} \cdot \nabla \phi$ the current generated by the electrochemical reaction at the corroding boundary. It is further assumed that the system remains well mixed at all times such that the gradients in species concentrations are negligible. Built upon that assumption, the current balance in the electrolyte is written as
\begin{equation} \label{eqn23}
    - \nabla \cdot \big(\xi \nabla \psi \big)  = c_0 F \sum R_i z_i + Q  \quad \text{in}\quad\Omega,
\end{equation}
where $\xi$ is the electric conductivity of the electrolyte whose magnitude depends on the amount of NaCl in the solution. The previous equation for the solution potential is solved in the entire computational domain. Yet, it has only physical meaning in the electrolyte phase. The governing equation for the solution potential is supplemented with two boundary conditions: $\xi \mathbf{n} \cdot \nabla \psi = 0$ on the boundary $\partial\Omega$ and $\psi = \psi_\mathrm{ref}$ on the boundary $\partial\Omega_\mathrm{ref}$, where $\psi_\mathrm{ref}$ stands for the solution potential on the reference electrode boundary $\partial\Omega_\mathrm{ref}$, Fig. \ref{Fig2}. Similar equations for the electric potential distribution can be found in Refs. \cite{Mai2018, Chadwick2018, Ansari2018, LIN2021, Cui2023, Makuch2024}.

\subsubsection{Mechanical equilibrium equation} \label{sec335}

The resulting set of governing equations is completed with the linear momentum balance equation for quasi-static loading and standard boundary conditions
\begin{equation} \label{eqn24}
\begin{aligned}
& \nabla \cdot \bm{\sigma} = \bm{0}  \quad\mathrm{}\quad \text{in } \Omega \\
\mathbf{t} = \mathbf{n}\cdot \bm{\sigma} = \mathbf{t}^0 \quad &\text{on}\quad \partial\Omega_{\mathbf{t}} \quad \text{and} \quad \mathbf{u} = \mathbf{u}^0 \quad\mathrm{} \text{on}\quad \partial\Omega_{\mathbf{u}},
\end{aligned}
\end{equation}
where $\mathbf{t}^0$ and $\mathbf{u}^0$ are the prescribed traction and displacement vectors. 

\subsubsection{Reaction rates} \label{sec346}

To complete the mathematical formulation, the reaction rates associated with the chemical reactions in Eqs. (\ref{eqn1a})$-$(\ref{eqn2b}) are defined as follows. The anodic reaction provides a source term for the evolution of $\text{Mg}^{2+}$ ions while the cathodic reaction generates hydrogen gas and $\text{OH}^{-}$ ions. Both reactions occur on the electrode surface and ascribe to the solid$-$liquid interface movement. The corresponding local reaction rates $R_i^{\prime}$ at the interface for each ionic species in the system are given as   
\begin{equation} \label{eqn25}
\begin{aligned}
R_1^{\prime} =- \frac{c_\mathrm{s}}{c_0} ( - \mathbf{v} \cdot \nabla \phi) = \frac{i}{c_0 F z}|\nabla \phi| \quad\mathrm{}\quad R_2^{\prime} = 2 R_1^{\prime} \quad\mathrm{}\quad R_3^{\prime} = 0.\\
\end{aligned}
\end{equation}
The expression for $R_1^{\prime}$ implies that the amount of dissolved Mg-metal due to the electrochemical reaction (convective transport in Eq. (\ref{eqn17})) is equivalent to the total amount of $\text{Mg}^{2+}$ ions generated on the surface of the metal electrode. The second equation for $R_2^{\prime}$ equates the amount of $\text{Mg}^{2+}$ and $\text{OH}^{-}$ ions produced at the solid$-$liquid interface according to Eqs. (\ref{eqn1a}) and (\ref{eqn1b}). $\text{Cl}^-$ ions are not involved in the electrochemical reaction at the electrode surface, and thus, their interface reaction rates are equal to zero. 

The volumetric reaction rate connected to the precipitation and dissolution of the corrosion product layer can be written as \cite{YAN1993}
\begin{equation} \label{eqn26}
\begin{aligned}
R_p^{\prime\prime} = k_p \big(\bar{c}_{1}\bar{c}_{2}^{2}-\bar{K}_\mathrm{sp}(\text{Cl}^-) \big) \quad\mathrm{}\quad R_p^{\prime\prime} = 0 \quad\text{if} \quad \bar{c}_{1}\bar{c}_{2}^{2} < \bar{K}_\mathrm{sp}(\text{Cl}^-),\\
\end{aligned}
\end{equation}
where $\bar{K}_\mathrm{sp} (\text{Cl}^-) = K_\mathrm{sp}(\text{Cl}^-)/c_0^3$ is the normalized solubility product constant of $\text{Mg(OH)}_2$ and $k_p$ the effective reaction rate responsible for the concurrent precipitation and dissolution of the Mg(OH)$_2$ layer. The condition in the previous equation ensures that the precipitation reaction only occurs after the saturation is reached. The role of Cl$^-$ ions in the precipitation of the corrosion product layer is included through the dependence of solubility product constant $K_\mathrm{sp}$ on chloride concentration (Section \ref{sec347}). $\text{Mg}^{2+}$ and $\text{OH}^-$ ions produced at the solid$-$liquid interface due to the electrochemical reaction (\ref{eqn25}) are consumed with the production of the corrosion product $\text{Mg(OH)}_2$ layer, Eq. (\ref{eqn1c}). At the same time, a certain amount of $\text{Mg}^{2+}$ and $\text{OH}^-$ ions are released to solution due to the dissolution of the layer, Eq. (\ref{eqn1d}). The consumption and release rates of these ions are controlled with the effective rate constant $k_p$. It is assumed here that MgCl$_2$ is unstable and instantaneously dissolves in $\text{Mg}^{2+}$ and 2Cl$^-$, Eq. (\ref{eqn1d}). This assumption eliminates tracking MgCl$_2$ as an independent concentration in the system. The amount of $\text{Cl}^-$ ions annihilated during the dissolution of the layer is immediately released in solution due to rapid decomposition of MgCl$_2$. The concentration of $\text{Cl}^-$ ions is thus unaltered under this assumption.

The volumetric reaction rates associated with the buffer system (Eqs. (\ref{eqn2a}) and (\ref{eqn2b})) are expressed as
\begin{equation} \label{eqn27}
\begin{aligned}
 R_\mathrm{pH}^{\prime\prime}  =  k_\mathrm{pH}(\bar{c}_2 - \bar{c}_2^\mathrm{lim})& \quad\mathrm{}\quad  R_\mathrm{pH}^{\prime\prime} = 0 \quad\text{if} \quad \bar{c}_2 < \bar{c}_2^\mathrm{lim}\\
 R_\mathrm{MgCO_3}^{\prime\prime} =  k_\mathrm{MgCO_3}(\bar{c}_1 - \bar{c}_1^\mathrm{lim})& \quad\mathrm{}\quad  R_\mathrm{MgCO_3}^{\prime\prime} = 0 \quad\text{if} \quad \bar{c}_1 < \bar{c}_1^\mathrm{lim}.\\
\end{aligned}
\end{equation}
$R_\mathrm{pH}^{\prime\prime}$ accounts for the consumption of $\text{OH}^{-}$ ions due to reaction (\ref{eqn2a}) and limits the pH of solution. $R_\mathrm{MgCO_3}^{\prime\prime}$ is responsible for the consumption of $\text{Mg}^{2+}$ ions due to the formation of MgCO$_3$, Eq. (\ref{eqn2b}). Here, $\bar{c}_2^\mathrm{lim}$ is the limit concentration of $\text{OH}^{-}$ ions for a desired pH level of the solution. $\bar{c}_1^\mathrm{lim} = c_\mathrm{MgCO_3} / c_0$ stands for the threshold concentration of $\text{Mg}^{2+}$ ions that must be reached to precipitate MgCO$_3$. This threshold is defined in terms of the molar solubility of MgCO$_3$, $c_\mathrm{MgCO_3} = 2.4 \times 10^{-4}$ M \cite{Thandi2007}. The condition ensures that the amount of $\text{Mg}^{2+}$ ions below the solubility limit remains in solution, whereas any excess in $\text{Mg}^{2+}$ ions beyond this limit is converted into MgCO$_3$. $k_\mathrm{pH}$ and $k_\mathrm{MgCO_3}$ serve as a penalty for departing from the limit concentrations $\bar{c}_1^\mathrm{lim}$ and $\bar{c}_2^\mathrm{lim}$. They must be large enough to enforce the equilibrium but not to obstruct the simulation. Combining Eqs. (\ref{eqn26}) and (\ref{eqn27}), the volumetric reaction rates for each ionic species are given as
\begin{equation} \label{eqn28}
\begin{aligned}
 R_1^{\prime \prime} = - R_p^{\prime\prime} - R_\mathrm{MgCO_3}^{\prime\prime} \quad\mathrm{}\quad R_2^{\prime \prime} = 2 R_p^{\prime \prime} - R_\mathrm{pH}^{\prime\prime} \quad\mathrm{}\quad R_3^{\prime \prime} = 0.\\
\end{aligned}
\end{equation}
Although there are no reaction rates associated with Cl$^-$ ions in the present model ($R_3^{\prime} = R_3^{\prime\prime} = 0$), they influence corrosion kinetics through the solubility product constant.

\subsubsection{Dependence of solubility product constant on the chloride concentration} \label{sec347}

\begin{figure}[h!]
    \centering
    \includegraphics[width = 16 cm]{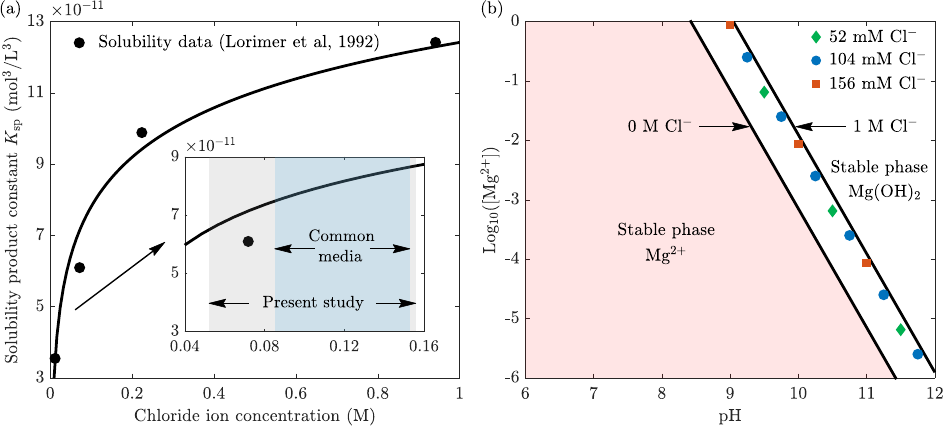}
    \captionsetup{labelfont = bf, justification = raggedright}
    \caption{Dependence of (a) solubility product constant $K_\mathrm{sp}$ and (b) chemical stability diagram on the concentration of chloride ions. The solubility experimental data in (a) is taken from Ref. \cite{Lorimer1992}. The light grey and blue areas stand for the range of chloride ion concentration used in the present study and in common corrosive media for Mg corrosion tests for biomedical applications \cite{MEI2020}. $[\text{Mg}^{2+}]$ denotes the molar concentration of Mg$^{2+}$ ions.}
    \label{Fig3}
\end{figure}

The solubility product constant $K_\mathrm{sp}$ determines the starting point of corrosion layer precipitation initiation. A lower value of the solubility product constant shortens the time required for the formation of the layer. In the present investigation, the dependence of $K_\mathrm{sp}$ on the chloride concentration is expressed by following experimental data on the solubility of Mg(OH)$_2$ in Cl$^-$ ions containing solutions \cite{Lorimer1992}
\begin{equation} \label{eqn29}
K_\mathrm{sp} (\text{Cl}^-) = K_\mathrm{sp}^{\text{1M}} + B\text{log}_{10}[\text{Cl}^-],
\end{equation}
where $K_\mathrm{sp}^{\text{1M}} = 1.24\times 10^{-10}$ mol$^3$/L$^3$ is the solubility product constant in 1 M Cl$^-$ and $B = 4.61\times10^{-11}$ mol$^3$/L$^3$ is a constant obtained through the fitting against experimental data \cite{Lorimer1992}. Here, $[\text{Cl}^{-}]$ denotes the molar concentration of Cl$^-$ ions. The thermodynamic equilibrium equation for the precipitation of the hydroxide layer (Eq. (\ref{eqn1c})) can be written as
\begin{equation} \label{eqn30}
\text{log}_{10} [\text{Mg}^{2+}] + 2 \text{log}_{10} [\text{OH}^{-}] =  \text{log}_{10} K_\mathrm{sp}(\text{Cl}^-) \quad\mathrm{}\quad \text{log}_{10} [\text{Mg}^{2+}] = 28 - 2 \text{pH} + \text{log}_{10} K_\mathrm{sp} (\text{Cl}^-),
\end{equation}
where $-\text{log}_{10} [\text{OH}^{-}] = 14 - \text{pH}$. $[\text{Mg}^{2+}]$ and $[\text{OH}^{-}]$ denote the molar concentration of Mg$^{2+}$ and OH$^{-}$ ions. The dependence of the solubility product constant and chemical stability diagram on the chloride concentration is depicted in Fig. \ref{Fig3}. The inset in Fig. \ref{Fig3}(a) shows the variation in the solubility product constant $K_\mathrm{sp}$ for the range of chlorine concentration considered in the current work and in common corrosive media used in Mg corrosion tests for biomedical applications \cite {MEI2020}. As indicated in Fig. \ref{Fig3}, a variation in chloride concentration from 52 mM Cl$^-$ to 156 mM Cl$^-$ has a negligible effect on the solubility product constant $K_\mathrm{sp}$ and the equilibrium condition for the precipitation of Mg(OH)$_2$ layer. This observation is experimentally validated in the current investigation, as discussed in the following section.  

\subsection{Dimensional analysis} \label{sec34}

The resulting set of governing equations is normalized using the interface thickness $\ell$ as the characteristic length, the diffusion coefficients of $\text{Mg}^{2+}$ ions $D_1$, and the energy barrier $\omega$ as the reference energy density. They render the following set of nondimensional governing equations
\begin{equation} \label{eqn31}
\left\{
\begin{aligned}
& \frac{\partial \phi}{\partial \bar{t}} = -\tau\Big(\frac{\partial \bar{\mathcal{F}}^\mathrm{chem}}{\partial \phi} + \frac{\partial \bar{\mathcal{F}}^\mathrm{mech}}{\partial \phi} -\bar{\kappa}\bar{\nabla}^2 \phi  \Big) - P_e |\bar{\nabla} \phi|\\
& \frac{\partial \bar{c}_i}{\partial \bar{t}} = \bar{\nabla} \cdot \Big[\bar{D}_i \big(\bar{\nabla} \bar{c}_i + z_i \bar{c}_i \bar{\nabla}\bar{\psi}\big)\Big] + \bar{R}_i^{\prime} + \bar{R}_i^{\prime\prime}\\
& -\bar{\nabla} \cdot  \big(\bar{\xi}_l \bar{\nabla} \bar{\psi}\big)  = \sum \bar{R}_i z_i + P_e |\bar{\nabla} \phi|\\
& \bar{\nabla} \cdot \bar{\bm{\sigma}} = \bm{0}
\end{aligned}
\right\}\
\quad\text{in }\Omega,
\end{equation}
along with the corresponding nondimensional boundary conditions. In the previous equation, $\bar{t} =tD_1/\ell^2$, $\bar{\mathbf{x}} = \mathbf{x}/\ell$, $\bar{\nabla} = \ell \nabla$, $\bar{\mathcal{F}}^\mathrm{chem} = \mathcal{F}^\mathrm{chem} / \omega$, $\bar{\mathcal{F}}^\mathrm{mech} = \mathcal{F}^\mathrm{mech} / \omega$, $\bar{\kappa} = \kappa / (\omega \ell^2)$, $\bar{D}_i = D_i / D_1$, $\bar{\psi} = \psi F / (RT)$, $\bar{\xi}_l = \xi_l R T / (D_1 c_0 F^2)$, and $\bar{\bm{\sigma}} = \bm{\sigma} / \omega$. The nondimensionalized reaction rates are defined as
\begin{equation} \label{eqn32}
\begin{aligned}
& \bar{R}_1^{\prime} =  \frac{c_s}{c_0} P_e |\bar{\nabla} \phi| \quad \bar{R}_2^{\prime} = 2\bar{R}_1^{\prime} \quad \bar{R}_3^{\prime} =  0\\
 \bar{R}_1^{\prime \prime} =  -  &\bar{R}_p^{\prime\prime} - \bar{R}_\mathrm{MgCO_3}^{\prime\prime} \quad \bar{R}_2^{\prime\prime} = 2 \bar{R}_p^{\prime \prime} - \bar{R}_\mathrm{pH}^{\prime\prime} \quad \bar{R}_3^{\prime\prime} = 0\\
 \bar{R}_p^{\prime\prime} = D_a (\bar{c}_{1}\bar{c}_{2}^{2}-\bar{K}_\mathrm{sp}&(\mathrm{Cl}^-))  \quad \bar{R}_\mathrm{pH}^{\prime\prime}  =  \bar{k}_\mathrm{pH}(\bar{c}_2 - \bar{c}_2^\mathrm{lim}) \quad \bar{R}_\mathrm{MgCO_3}^{\prime\prime} =  \bar{k}_\mathrm{MgCO_3}(\bar{c}_1 - \bar{c}_1^\mathrm{lim}).
\end{aligned}
\end{equation}

The characteristic times for diffusion of ions within the bulk solution $t_d$, interface migration due to the interfacial energy reduction $t_\phi$, interface motion due to the electrochemical reaction (convective term in Eq. (\ref{eqn12})) $t_v$, volumetric reaction rate for the precipitation/dissolution of the corrosion product layer $t_{k_p}$ are
\begin{equation} \label{eqn33}
                   t_d =\frac{\ell^2}{D^l_1} \quad\mathrm{}\quad t_\phi = \frac{1}{L \omega} \quad\mathrm{}\quad t_v = \frac{\ell}{v} \quad\mathrm{}\quad t_{k_p} = \frac{1}{k_p},
\end{equation}
where $v = | \mathbf{v}| = i /(c_s F z)$ is the magnitude of the interface velocity due to the electrochemical reaction (\ref{eqn13}). The relative strength of convection, diffusion, and reactions is represented by the following dimensionless numbers
\begin{equation} \label{eqn34}
                   \tau_\phi =\frac{t_v}{t_\phi} \quad\mathrm{}\quad P_e = \frac{t_d}{t_v} \quad\mathrm{}\quad \tau =\frac{t_d}{t_\phi} = P_e\tau_\phi \quad\mathrm{}\quad D_a = \frac{t_d}{t_{k_p}}. 
\end{equation}
The rate of material transport at the interface is controlled by $\tau_\phi$, $\tau$, and the Peclet number $P_e$ while the effective reaction kinetics for the formation/dissolution of the corrosion product layer is governed by the Damk\"ohler number $D_a$. The interface migration is dominantly governed by the electrochemical reaction at the solid$-$liquid interface compared to the diffusion part in Eq. (\ref{eqn12}). $\tau_\phi$ is the ratio between convective and diffusive transport of the phase-field variable. As the interface is governed by the electrochemical reaction, the condition $\tau_\phi \ll 1$ provides the criterion for the interfacial mobility coefficient that reads $L \ll v/(\ell \omega)$. The ratio between bulk diffusion and interface diffusion is controlled by $\tau$. As the characteristic time for interface diffusion is large, it is expected that $\tau = P_e \tau_\phi \ll 1$.

The Peclet number $P_e$ determines the rate-limiting process for material transport between bulk diffusion and the electrochemical reaction at the interface. When $P_e\gg 1$, the electrochemical reaction is much faster than the diffusion and the process is driven by bulk diffusion (diffusion-controlled corrosion). For $P_e \ll 1$, diffusion is faster than the electrochemical reaction such that there is no accumulation of species at the solid$-$liquid interface (activation-controlled corrosion). Two extreme scenarios are possible regarding the volumetric reaction kinetics ascribed to the formation/dissolution of the Mg(OH)$_2$ layer. $D_a = 0$ implies that the layer does not form (the formation and dissolution rates are equal) and that all Mg$^{2+}$ and OH$^-$ ions generated in (\ref{eqn25}) stay within the electrolyte. For $D_a \gg 0$ the formation rate dominates and the layer nucleates. $D_a < 0$ is not physical as it would imply that dissolution occurs faster than formation. Hence, $D_a \gg 0$ is enforced in this investigation and its value is determined through the fitting against the experimental measurements on the concentration of Mg$^{2+}$ ions in the solution. $\bar{k}_\mathrm{pH}$ and $\bar{k}_\mathrm{MgCO_3}$ in Eq. (\ref{eqn32}) are the normalized penalty coefficients that enforce the limit concentrations associated with the buffer system.

\section{Model calibration and validation}\label{sec4}

The model is calibrated and validated against the experimental measurements of mass loss, the average concentration of Mg ions in solution, and pH over the immersion time. The material properties of the metal and the diffusivity of ions used in the simulation are listed in Table \ref{table1}. For consistency with experimental observations, the standard electrode equilibrium potential $E^{\theta}$ and the electric conductivity of the electrolyte $\xi$ are made dependent on chloride concentrations. $E^{\theta}$ is estimated from potentiodynamic polarization curves \cite{Rettig2008, ACHARYA2019, GUADARRAMAMUNOZ2006} and is set to $E^{\Theta} = -1.20 $ V (vs. SHE) for the reference solution of 104 mM NaCl. For the 52 mM NaCl and 156 mM NaCl solutions, the values are set to $E^{\Theta} = -1.1975$ V (vs. SHE) and $E^{\Theta} = -1.2025$ V (vs. SHE) following reported variations in the literature \cite{ZHAO2008, Cain2014, Wang2010}. The electric conductivity of the electrolyte is set to $\xi = 0.52$ S/m for 52 mM NaCl, $\xi = 1$ S/m for 104 mM NaCl, and $\xi = 1.375$ S/m for 156 mM NaCl \cite{HALLSTEDT2007}. The anodic charge transfer coefficient $\alpha_{\mathrm{a}}$ is estimated from potentiodynamic polarization curves \cite{Rettig2008, ACHARYA2019, GUADARRAMAMUNOZ2006}. It is assumed to be independent of chloride concentration and set to $\alpha_{\mathrm{a}} = 0.30$. The normalized solubility product constant $\bar{K}_\mathrm{sp}$ is defined using Eq. (\ref{eqn29}). The charge numbers of ionic species considered $z_i$ have common values, as stated in Eqs. (\ref{eqn1a})$-$(\ref{eqn1d}). The phase-field model parameters $\omega$ (Eq. (\ref{eqn4})) and $\kappa$ (Eq. (\ref{eqn5})) are related to the interfacial energy $\Gamma$ and the nominal chosen interface thickness $\ell$. The interface thickness is set to be significantly smaller than the characteristic domain size ($\ell$ = 5 $\upmu$m). The simulation for model calibration and validation assumes uniform corrosion and the mechanical effect (Eq. (\ref{eqn16})) is not considered.

Three additional parameters need to be specified: the exchange current density $i_0$, the reaction rate for the formation/dissolution of the corrosion product layer $D_a$, and the phase-field mobility parameter $L$ in Eq. (\ref{eqn17}). The exchange current density is a physical parameter that depends on the pH and temperature of the solution. Due to a lack of experimental data on this parameter in NaCl solutions and at the temperature considered, the exchange current density is determined by fitting against experimental measurements. The same fitting approach is applied to the Damk\"ohler number $D_a$. The phase-field mobility parameter $L$ is selected based on dimensional analysis (Section \ref{sec34}) such that the condition $L \ll v/(\ell \omega)$ is satisfied. $L = v/(10 \ell \omega)$ is used in the present investigation. These three parameters are the same for the air and 5\% CO$_2$ environments. To enforce the buffer effects, the penalty coefficients $\bar{k}_\mathrm{pH}$ and $\bar{k}_\mathrm{MgCO_3}$ need to be defined. They should be chosen large enough to enforce equilibrium conditions (\ref{eqn32}). While larger values achieve equilibrium more quickly, they can significantly increase computational costs by reducing the time step and potentially cause convergence issues. Hence, they should be selected judiciously. In the present study, the penalty coefficients are set to $\bar{k}_\mathrm{pH} = 1$ and $\bar{k}_\mathrm{MgCO_3} = 10$ as a trade-off between accuracy and computational efficiency. Under the assumed condition for $L$ and adopted anodic charge transfer and penalty coefficients, the current model has only two parameters that need to be fitted against experimental data: the exchange current density $i_0$ and the Damk\"ohler number $D_a$ for both buffer-free and buffer-controlled media. 

\begin{table}[t]
\centering
\begin{tabular}{ l l l } 
 \hline
 Quantity & Value & Unit\\
\hline
Absolute temperature $T$ & 310.15 & K\\
Molar volume of magnesium $V_\mathrm{m}$ & 13.998 & cm$^3$/mol \cite{HALLSTEDT2007}\\
Diffusion coefficient of $\text{Mg}^{2+}$ ions $D_1$ & 7.06$\times 10^{-10}$ & m$^2$/s \cite{Haynes2016}\\
Diffusion coefficient of $\text{OH}^{-}$ ions $D_2$ & 5.273$\times 10 ^{-9}$ & m$^2$/s \cite{Haynes2016}\\
Diffusion coefficient of $\text{Cl}^{-}$ ions $D_3$ & 2.032$\times 10 ^{-9}$ & m$^2$/s \cite{Haynes2016}\\
Standard concentration in the electrolyte $c_0$ & 1 & mol/L\\
Interfacial energy $\Gamma$ & 0.5 & J/m$^2$ \cite{FU2009, Skriver1992}\\
Anodic charge transfer coefficient $\alpha_\mathrm{a}$ & 0.30 & - \cite{Rettig2008, ACHARYA2019, GUADARRAMAMUNOZ2006}\\
\hline
\end{tabular}
\captionsetup{labelfont = bf,justification = centering}
\caption{Parameters used in all phase-field simulations.}
\label{table1}
\end{table}

\begin{figure}[h!]
    \centering
    \includegraphics[width = 16 cm]{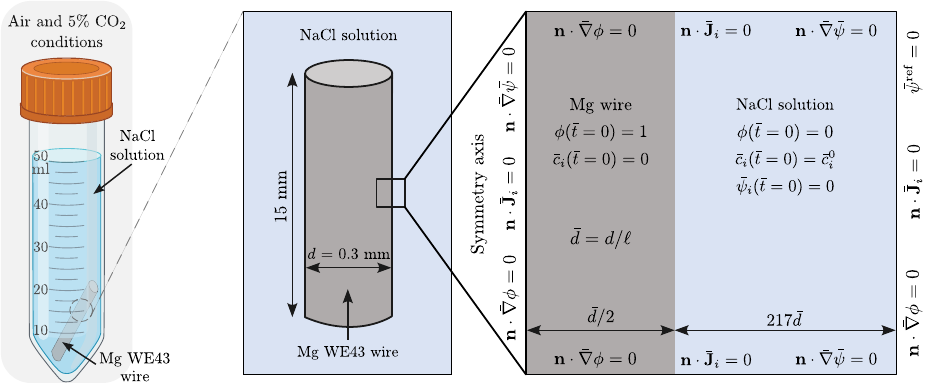}
    \captionsetup{labelfont = bf, justification = raggedright}
    \caption{Schematic representation of the experimental setup (left) and the corresponding nondimensional computational domain (right) for the Mg alloy wire immersed in different NaCl solutions.}
    \label{Fig4}
\end{figure}

The numerical simulation is conducted by considering a one-dimensional axisymmetric domain as depicted in Fig. \ref{Fig4}. The actual diameter of the wire utilized in the experiments, $d = 0.30$ mm, is considered in the simulation. To accurately reflect the experimental conditions, the size of the electrolyte is determined by preserving the volume ratio between the Mg alloy and the amount of solution used in the degradation tests. Using the dimensions of the wire, a total of 50 mL of solution, and assuming the electrolyte has a cylindrical shape, it is found that the equivalent radius of the electrolyte is approximately $217  d$, Fig. \ref{Fig4}. The nondimensional form of governing equations (\ref{eqn31}) is solved with accompanying initial and boundary conditions for each primary field variable. The initial concentration values of each ionic species are based on the experimental setup. The initial concentration of Mg$^{2+}$ ions in the domain is set to zero. A pH of 5.5 is used to calculate the initial concentrations of OH$^-$ ions. The initial concentration of Cl$^-$ ions in the medium matches the values used in the experiment. A zero-flux boundary condition for the diffusion of all ionic species and phase-field variable is enforced at the outer boundaries of the domain to resemble the closed environment of the \textit{in vitro} experiments. This boundary condition ensures that no diffusion occurs across the domain boundary. A zero reference solution potential is imposed on the far right boundary in the electrolyte. The remaining boundaries are treated as perfectly isolated with no flux boundary condition for the solution potential.

\begin{figure}[h!]
    \centering
    \includegraphics[width = 16 cm]{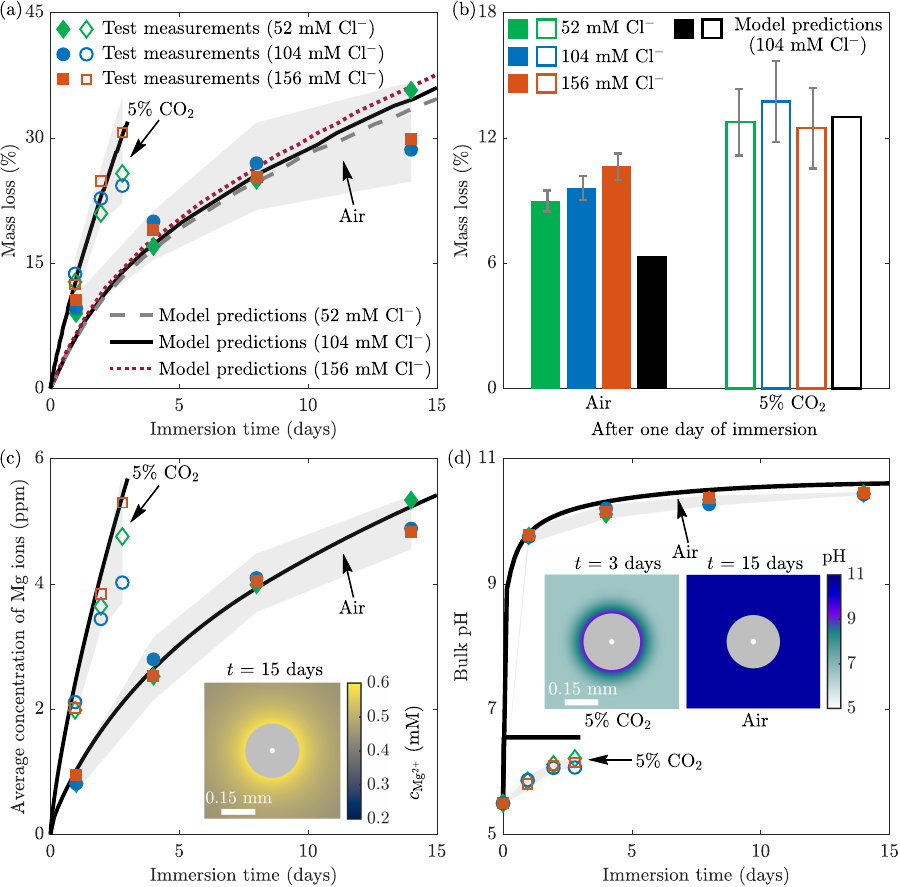}
    \captionsetup{labelfont = bf, justification = raggedright}
    \caption{Comparison between experimental measurements and phase-field predictions for (a) mass loss, (b) mass loss after one day of immersion, (c) average concentration of Mg ions in solution, and (d) bulk pH. The light grey area stands for the standard deviation of the experiments considering data of all three NaCl solutions. The legends in the insets in (c) and (d) apply to the whole computational domain. The white point and grey area in the insets in (c) and (d) stand for the center and final cross-section of the Mg wire after degradation.}
    \label{Fig5}
\end{figure}

\begin{figure}[h!]
    \centering
    \includegraphics[width = 16 cm]{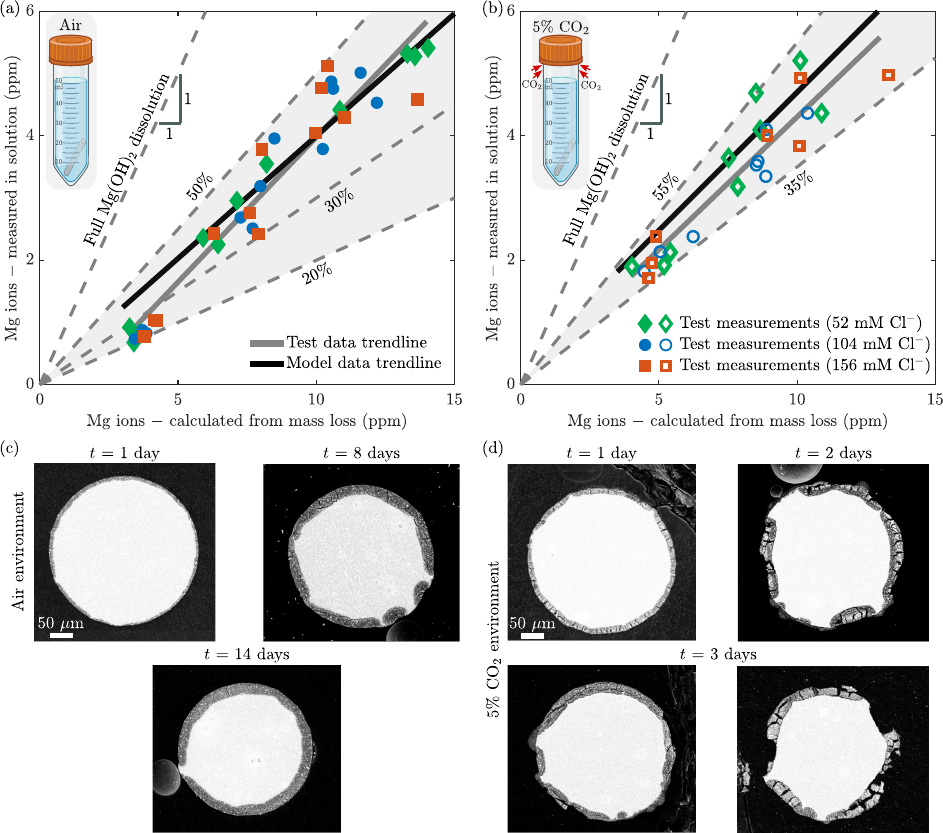}
    \captionsetup{labelfont = bf, justification = raggedright}
    \caption{Correlation between the concentration of Mg ions computed based on mass loss measurements and the detected concentration of Mg ions in solution for model predictions and tests in (a) air and (b) 5\% CO$_2$. Scanning electron microscopy images of cross-sections of the wire at distinct times of degradation for tests in (c) air and (d) 5\% CO$_2$. Mg appears in white while the Mg(OH)$_2$ layer is gray.}
    \label{Fig6}
\end{figure}

The predicted mass loss, the average concentration of Mg ions in solution, and pH levels are plotted against immersion time in Fig. \ref{Fig5}, together with the experimental measurements obtained from the \textit{in vitro} tests. The phase-field predictions and the test measurements are in good agreement for both air and 5\% CO$_2$ environments. The qualitative difference in mass loss (Fig. \ref{Fig5}(a) and Fig. \ref{Fig5}(b)) and the concentration of Mg ions in solution (Fig. \ref{Fig5}(c)) between these two test environments yields one consistent result: the corrosion rate and the quantity of Mg ions in solution are higher in the buffer solution than in the air environment. Moreover, the experimental results conclusively demonstrated that a variation in chloride concentration from 52 mM to 156 mM had a negligible effect on mass loss, the amount of released Mg$^{2+}$ ions in solution, and pH levels. This finding confirms that the variation in Cl$^-$ ions considered does not affect the solubility limit of Mg(OH)$_2$, as indicated in Fig. \ref{Fig3}. The consistent increase in mass loss correlated with a rise in Cl$^-$ ions is only noticeable after the first day of immersion in the air environment, Fig. \ref{Fig5}(b). After that, and in the case of the 5\% CO$_2$ environment, the experimental measurements show no correlation with the amount of Cl$^-$ ions. This conclusion shows that variations in pH are the likely cause of differences in Mg corrosion rates rather than fluctuations in Cl$^-$ ion concentration in the solution. The experiments indicated that the corrosion rate was initially high in the three NaCl solutions within the air environment. However, this rate slowed down over time and eventually reached a steady rate. The initial rapid corrosion rate was attributed experimentally to a low pH value (Fig. \ref{Fig5}(d)), which was well below the threshold required for forming the corrosion product layer, Fig. \ref{Fig3}(b). The progress of corrosion after the first day of immersion led to an increase in Mg$^{2+}$ ions and pH, Fig. \ref{Fig5}(c) and Fig. \ref{Fig5}(d). The pH increased abruptly and then slowly increased with further immersion time, reaching a steady state value of 10.5, Fig. \ref{Fig5}(d). These favorable conditions in terms of high Mg$^{2+}$ ions and elevated pH initiated the nucleation of the Mg(OH)$_2$ layer (Fig. \ref{Fig3}(b)), which subsequently reduced the corrosion rate. The pH also rapidly increased in the 5\% CO$_2$ environment and stayed at the target value of around 6.2. This low pH value undermined the mechanical integrity of the Mg(OH)$_2$ layer, resulting in significantly higher mass loss and concentration of Mg$^{2+}$ ions in the solution from the beginning of the immersion period. SEM images of the Mg(OH)$_2$ layer at distinct time intervals in both test environments are provided in Fig. \ref{Fig6} and discussed further below.  

The phase-field predictions capture the experimental trends and accurately reproduce the measurements of mass loss, the average concentration of Mg ion in solution, and pH values for both air and 5\% CO$_2$ environments, Fig. \ref{Fig5}. The mass loss predictions for the three NaCl solutions in the air environment are presented in Fig. \ref{Fig5}(a). Since the range of Cl$^-$ ion concentrations considered in the present work does not influence the corrosion of the Mg wire, the predictions for the average concentration of Mg ions in solution and pH are only provided for the reference solution containing 104 mM NaCl. As can be observed from Fig. \ref{Fig5}(a) and Fig. \ref{Fig5}(b), the current framework slightly underestimates mass loss measurements in the air environment at very early immersion times. It tends to slightly overestimate the bulk pH value in the 5\% CO$_2$ environment, as illustrated in Fig. \ref{Fig5}(d). The model precisely resembles the concentration of Mg ions in solution for both air and 5\% CO$_2$ environments, as shown in \ref{Fig5}(c). The inset in Fig. \ref{Fig5}(c) portrays the distribution of Mg ions within the solution for the air environment at the final computational time. Two additional insets in Fig. \ref{Fig5}(d) show the pH distribution in both environments at the final computational times. These insets indicate that a constant pH value of 10.5 is maintained throughout the domain in the air environment at the end of immersion. Albeit the bulk pH value is constrained in the 5\% CO$_2$ environment, a nonuniform pH distribution is observed, with a significantly higher pH (around 9.2) near the interfacial area compared to the bulk solution. This significant difference in local (near wire surface) and bulk pH is physical and it has been observed before \cite{SONG1997_pH, Lamaka2018}. A quantitative comparison is performed between experimental data and model predictions considering two metrics: mean absolute error (MAE) and root-mean-squared error (RMSE). The MAE values of 3.30\%, 0.13, and 0.19 ppm are computed for mass loss, pH, and concentration of Mg$^{2+}$ ions in the air environment. The corresponding values of 2.67\%, 0.55, and 0.92 ppm are returned in the 5\% CO$_2$ environment. The RMSE values between experimental and predicted values for mass loss, pH, and concentration of Mg$^{2+}$ ions are 3.65\%, 0.15, and 0.21 ppm in the air environment. The corresponding RMSE values of 4.27\%, 0.56, and 1.07 ppm are obtained in the 5\% CO$_2$ environment. The reported MAE and RMSE values are determined for the reference solution containing 104 mM NaCl. The model returns the experimental data using an exchange current density $i_0 = 7.25 \times 10^{-11}$ A/cm$^2$ and a Damk\"ohler number $D_a = 2500$. The returned exchange current density falls within the reported range of $10^{-6}$ A/cm$^2$ to 10$^{-12}$ A/cm$^2$ \cite{SONG1997, FAJARDO2015, FRANKEL20131}. The agreement between the model predictions and the experimental data on the concentration of Mg$^{2+}$ ions in solution and the pH verifies that the Damk\"ohler number is adequately chosen to accurately produce experimental measurements. The match with the experimental results for the concentration of Mg ions in solution and pH values in the 5\% CO$_2$ environment indicates that the penalty coefficients $\bar{k}_\mathrm{pH}$ and $\bar{k}_\mathrm{MgCO_3}$ are suitably selected. A further increase in these coefficients would bring the model predictions closer to the experimental data at the cost of smaller time steps and longer computational times.

The dissolution and stability of the corrosion product Mg(OH)$_2$ layer under the two environments considered are analyzed using data on mass loss and the concentration of Mg ions in the solution. Fig. \ref{Fig6} illustrates the correlation between the concentration of Mg ions calculated from mass loss and the concentration measured in solution for both air and 5\% CO$_2$ environments. The concentration derived from mass loss assumes that all Mg ions resulting from mass loss are released into the solution, implying that these ions do not contribute to the corrosion product layer. As portrayed in Fig. \ref{Fig6}, the model and experimental data trendlines are in good agreement. In the air environment, the analysis reveals that 20-50\% of the Mg ions generated from mass loss are found in the solution, while the remainder stays within the Mg(OH)$_2$ layer. Excluding the initial rapid corrosion rate after the first day of immersion (Fig. \ref{Fig5}(b)), the concentration of Mg ions in the electrolyte increases to 30-50\%, Fig. \ref{Fig6}(a). This finding signifies that more Mg ions are consumed for the precipitation of the Mg(OH)$_2$ layer during the early stages of immersion than during longer immersion times. The phase-field data trendline falls within this 30-50\% range, consistent with Fig. \ref{Fig5}(b), as the model slightly underestimates mass loss after one day of immersion. In the case of the 5\% CO$_2$ environment, the model and experimental data trendlines range from 35-55\%, implying that a larger portion of the Mg(OH)$_2$ layer dissolves into solution, Fig. \ref{Fig6}(b). This suggests that the Mg(OH)$_2$ layer is less stable in the buffer-controlled environment with a lower pH value. The scanning electron microscopy images of the corrosion product layer at different time intervals for both test environments are shown in Fig. \ref{Fig6}(c) and Fig. \ref{Fig6}(d). As indicated in Fig. \ref{Fig6}(c) for the air environment, the layer is porous and contains microcracks. The layer maintains its compact structure throughout the immersion period, reducing the corrosion rate, Fig. \ref{Fig5}(a). In the case of the 5\% CO$_2$ environment, the layer is more porous with larger cracks present even at the early stage of immersion, Fig. \ref{Fig6}(d). The compactness of the layer diminishes as the cracks widen, facilitating fast ion diffusion through it. This dispersion of the layer initiates severe localized corrosion on the alloy surface in the early stage of immersion. As immersion time continues to increase, the cracks begin to merge, causing the detachment of large portions of the layer and exposing fresh Mg alloy to the corrosive environment, leading to high corrosion rates, Fig. \ref{Fig5}(a). The differences in corrosion behavior between the air and 5\% CO$_2$ environments are owing to the carbonate-bicarbonate pH buffering effects, Eqs. (\ref{eqn2a}) and (\ref{eqn2b}). The current \textit{in vitro} tests and phase-field predictions show that the WE43MEO Mg alloy exhibits better corrosion resistance when tested in air than in a buffer-regulated solution. 

\section{Implications} \label{sec5}

The present model is used in this section to predict the evolution of corrosion in the presence of mechanical loading in two different practical scenarios for orthopedic applications. The first case study analyzes bioabsorbable Mg alloy plates and screws for bone fracture fixation. The second one examines the corrosion of bioabsorbable Mg alloy porous scaffolds for the healing of critical-sized bone defects.

\subsection{Bioabsorbable Mg alloy plates and screws for bone fracture fixation} \label{sec51}

\begin{figure}[h!]
    \centering
    \includegraphics[width = 16 cm]{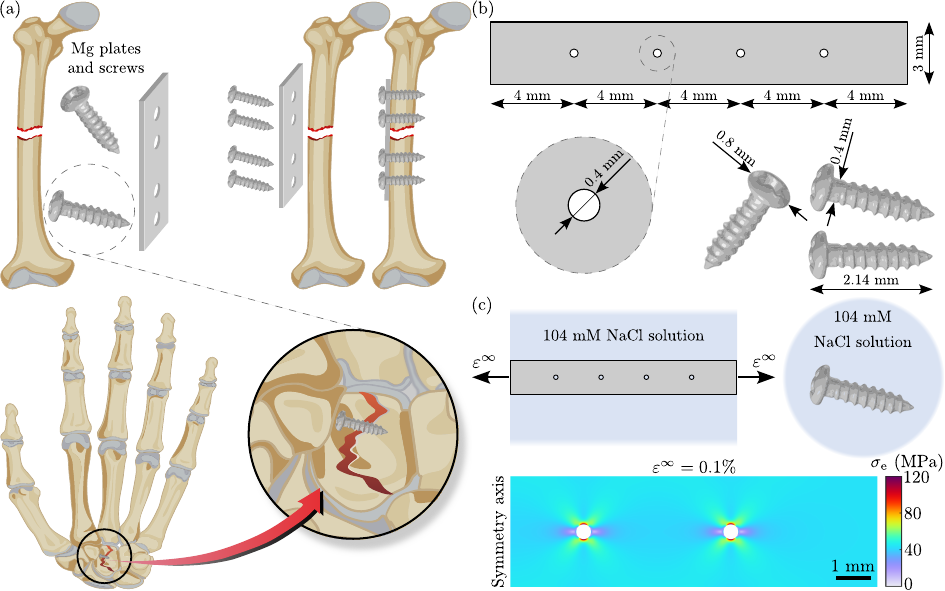}
    \captionsetup{labelfont = bf, justification = raggedright}
    \caption{Bioabsorbable Mg alloy plates and screws for bone fracture fixation. (a) Representative examples of bone fracture fixation. (b) The geometry and size of the Mg plate and screw considered in the simulation. (c) Corresponding computational domains and corrosive environment for the Mg plate and screw. Distribution of von Mises stress $\sigma_{\mathrm{e}}$ for the Mg plate prior to implantation for the remote tensile strain $\varepsilon^\infty = 0.1$\%.}
    \label{Fig7}
\end{figure}

Bioabsorbable Mg alloys are appealing for bone fracture fixation and internal fixation devices, as depicted in Fig. \ref{Fig7}(a), that help facilitate healing \cite {BAIRAGI2022}. During degradation, they stimulate new bone formation and stabilize uninhibited healing. Rapid degradation rates lead to premature implant failures that require additional surgery for implant replacement. The performance of these implants is determined in \textit{in vivo} and \textit{in vitro} tests \cite{KRAUS2012, CHAYA2015}. This study demonstrates that the corrosion performance of bioabsorbable implants for bone fracture fixation can be effectively assessed utilizing the current model.

A Mg alloy flat plate with 20 mm in length, 3 mm in width, and 1 mm in thickness is considered in the simulation. The plate is attached to the fractured bone using four Mg alloy screws, as schematically shown in Fig. \ref{Fig7}(b). The pre-drilled holes in the plate are 0.40 mm in diameter. The distance between the holes in the plate is 4 mm. The screws are 2.14 mm in length with a shaft outer diameter of 0.80 mm and a shaft inner diameter of 0.40 mm, Fig. \ref{Fig7}(b). The chosen Mg plate and screws are commonly referenced in the literature \cite{CHAYA2015, ZHAO2017}. The size of the corrosive medium is significantly larger than the size of the Mg plate and the screws. It is assumed in the present investigation that the plate and the screws are manufactured from the same Mg alloy. The material properties follow those used in the previous section and Table \ref{table1}. The phase-field model parameters $\omega$ and $\kappa$ are expressed using a nominal chosen interface thickness of $\ell$ = 20 $\upmu$m, which is considerably smaller than the characteristic sizes of the plate and screws. 

\begin{figure}[h!]
    \centering
    \includegraphics[width = 16 cm]{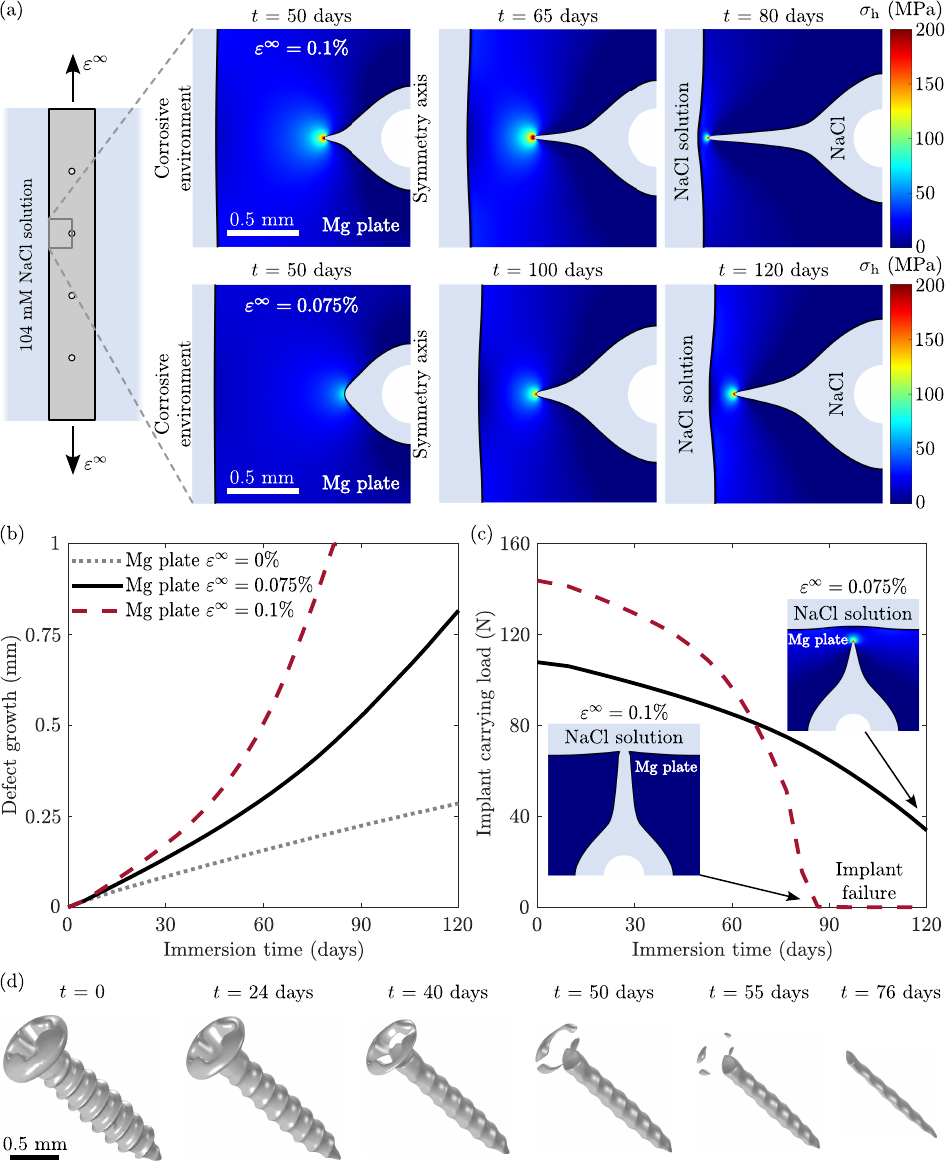}
    \captionsetup{labelfont = bf, justification = raggedright}
    \caption{Bioabsorbable Mg alloy plates and screws for bone fracture fixation. (a) Evolution of the metal$-$electrolyte interface and distribution of the hydrostatic stress, (b) defect growth, and (c) implant carrying load for the Mg plate subjected to various remote tensile strain $\varepsilon^\infty$. (d) Evolution of degradation of the Mg screw as a function of immersion time.}
    \label{Fig8}
\end{figure}

The Mg plate and screw are immersed in a NaCl-based solution containing 104 mM of Cl$^-$. The initial concentration of Mg$^{2+}$ is set to 0.875 mmol/L. A pH of 7.4 is used to compute the initial concentration of OH$^-$ ions. These initial conditions for the ionic species are chosen to replicate the conditions of human blood. Without loss of generality, the pH is not controlled in this example. As such, the penalty coefficients $\bar{k}_\mathrm{pH}$ and $\bar{k}_\mathrm{MgCO_3}$ are set to zero. This choice is made primarily to reduce computational costs, as the timescale of degradation is significantly longer, as illustrated in Fig. \ref{Fig8}, compared to the previous example with Mg wires. As discussed in the previous section, applying the penalty coefficients to regulate the pH would lower the computational time step. Hence, this decision should be viewed as a trade-off to demonstrate the potential of the model, rather than a limitation. Two distinct sets of simulations are carried out. In the first set, only the Mg alloy plate is simulated without the screws. In the second set, only the screws are analyzed, neglecting the Mg plate. This approach is adopted for simplicity as it additionally reduces computational costs and enables to simulate only one screw. Simulating only the screw is practical for the fixation of small bones and bone fragments where screws serve as non-load-bearing implants, Fig. \ref{Fig7}(a). This simplification allows for 2D simulations of the Mg plate and 3D simulations of the screw. The boundary conditions for the phase-field variable, the concentration of ionic species, and the solution potential are the same as in Section \ref{sec4}.

The role of mechanical fields in degradation is only considered for the Mg plate. The mechanical properties of the material are based on the uniaxial tensile tests conducted on the WE43MEO wires (Section \ref{sec2}). Lame's constants are set to $\lambda = 26.25$ GPa and $\mu = 17.50$ GPa. The hardening exponent $N$ in Eq. (\ref{eqn9}) is set to $N = 0.09$. The onset of strain for hardening in Eq. (\ref{eqn9}) is defined as $\epsilon_0 = \sigma_y / E$, where $E = 45.50$ GPa is Young's modulus. It is assumed in the simulation that the plate is subjected to a constant remote tensile strain $\varepsilon^\infty$. Three different magnitudes of the tensile loading are considered to show the sensitivity of degradation to mechanical fields: $\varepsilon^\infty = 0$, $\varepsilon^\infty = 0.075$\%, and $\varepsilon^\infty = 0.1$\%. The highest load is chosen so that the maximum stress prior to implantation is well below the yield strength $\sigma_y$. The loading is applied at the far right and left boundaries, Fig. \ref{Fig7}(c). The load is kept constant throughout the simulation. A quarter of the Mg plate is used in the simulation due to the symmetry conditions. In addition to the boundary conditions described above, an additional constraint is required for the mechanical equilibrium equation (\ref{eqn24}). The normal component of the displacement vector is constrained along the horizontal and vertical axes of symmetry, $\mathbf{n} \cdot \mathbf{u} = 0$. It is assumed that the boundaries where the load $\varepsilon^\infty$ is prescribed do not corrode. All other external boundaries and pre-drilled holes are exposed to the corrosive medium. The distribution of von Mises stresses at the initial computational time is illustrated in Fig. \ref{Fig7}(c). Mechanical loading is not applied to the screws as they are not included in the simulation with the plate. Consequently, this simulation for the screw represents uniform corrosion and is conducted to compare its degradation kinetics with those of the plate.

\begin{figure}[h!]
    \centering
    \includegraphics[width = 16 cm]{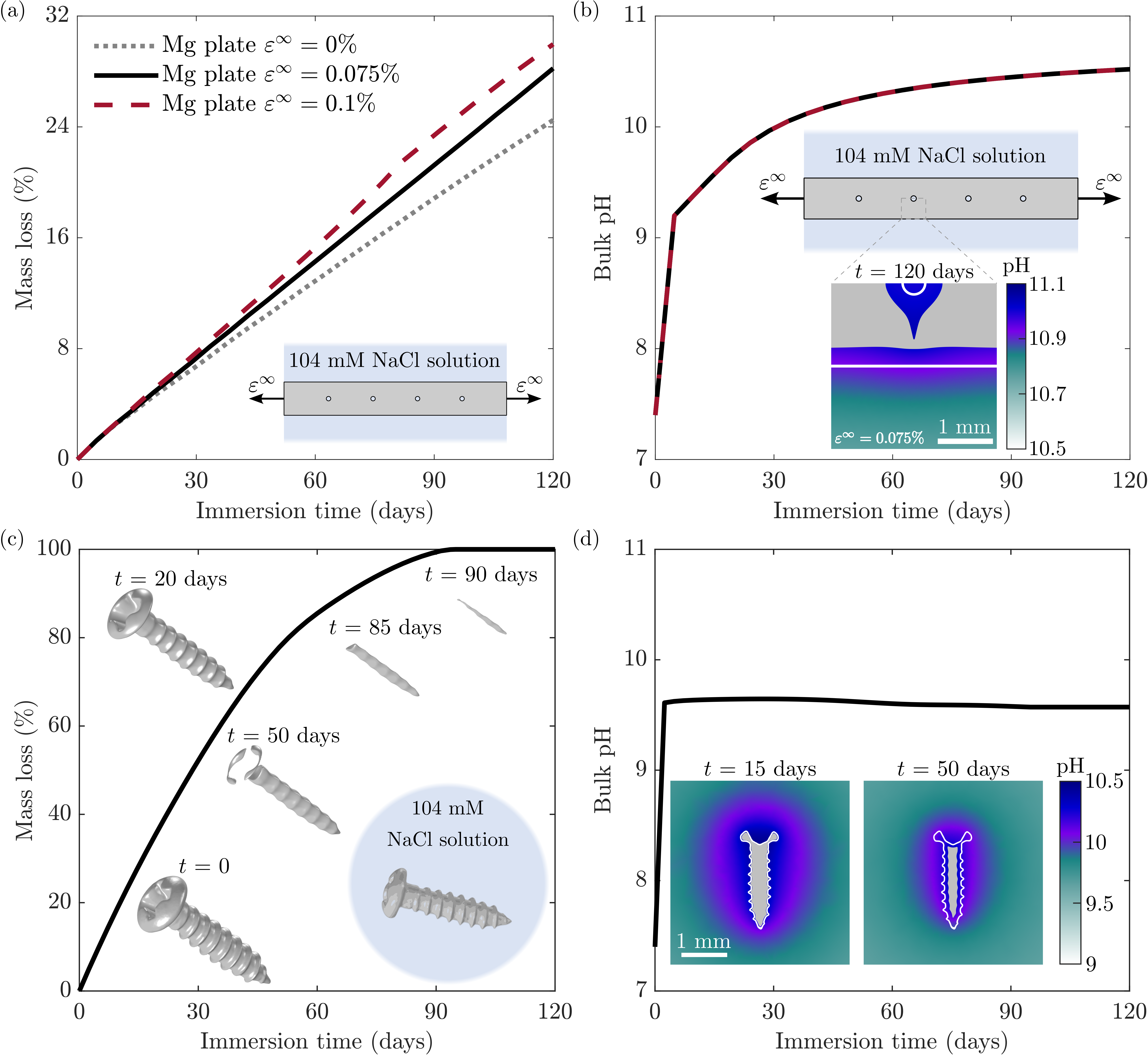}
    \captionsetup{labelfont = bf, justification = raggedright}
    \caption{Bioabsorbable Mg alloy plates and screws for bone fracture fixation. (a) Mass loss and (b) bulk pH as a function of immersion time for the Mg plate subjected to various remote tensile strain $\varepsilon^\infty$. (c) Mass loss and (d) bulk pH as a function of immersion time for the Mg screw. The white line in the insets in (b) and (d) indicates the initial metal$-$electrolyte interface. The legends in the insets in (b) and (d) apply to the whole computational domain.}
    \label{Fig9}
\end{figure}

Fig. \ref{Fig8} displays the phase-field results related to the performance of the Mg alloy plate as a function of immersion time and loading conditions $\varepsilon^\infty$. All simulations are performed for an immersion period of up to one hundred twenty days. The degradation of the Mg plate occurs simultaneously from both the exposed external boundaries and the pre-drilled holes. The evolution of the metal$-$liquid interface and the distribution of hydrostatic stress as a function of immersion time are illustrated in Fig. \ref{Fig8}(a). As indicated in Fig. \ref{Fig8}(a), more pronounced dissolution effects are observed around the pre-drilled holes where mechanical fields dominate compared to the external boundaries, which exhibit a constant corrosion rate. Mechanical fields lower the equilibrium electrode$-$electrolyte potential difference, thereby increasing the overpotential $\eta$ in Eq. (\ref{eqn14}). As a result, dissolution rates accelerate locally in areas with stresses and plastic strains. A higher increase in the overpotential is correlated with higher mechanical loading. The locally intensified degradation rate has a negligible impact on mass loss and bulk pH, Fig. \ref{Fig9}(a) and Fig. \ref{Fig9}(b). However, the locally enhanced degradation rate has a tremendous effect on the defect growth and implant carrying load, Fig. \ref{Fig8}(b) and Fig. \ref{Fig8}(c). Herein, the defect growth represents the evolution of the corrosion front from the pre-drilled holes toward the external longitudinal boundaries.

The presence of mechanical loading promotes the formation of cracks (Fig. \ref{Fig8}(a)), ultimately leading to the rapid fracture of the implant, Fig. \ref{Fig8}(b). Fig. \ref{Fig8}(a) shows that sharp cracks form after sixty-five days of immersion with an external loading of $\varepsilon^\infty = 0.1$\%. The results also indicate that the applied mechanical loading of $\varepsilon^\infty = 0.1$\% causes complete implant rupture after ninety days of immersion, see the inset in Fig. \ref{Fig8}(c). It is emphasized that negligible plastic strains occur just before implant failure, and thus, their distribution is not included here for brevity. No plastic strain develops in the case of $\varepsilon^\infty = 0.075$\% after three months of immersion. This study demonstrates that the Mg plate under a strain of $\varepsilon^\infty = 0.1$\% can be used for fast bone healing cases, as the implant strength decreases quickly. For instance, mechanical strength is reduced by 50\% after two months of immersion. In the case of $\varepsilon^\infty = 0.075$\%, the carrying capacity of the plate drops slowly and the implant maintains its mechanical integrity even after three months of immersion. The critical cross-section where fractures occur lies between the pre-drilled hole and the longitudinal external boundaries of the implant.

The corrosion performance of the Mg alloy screw in terms of change in volume as a result of corrosion is illustrated in Fig. \ref{Fig8}(d). The plots for mass loss and bulk pH are shown in Fig. \ref{Fig9}(c) and Fig. \ref{Fig9}(d). As can be seen from Fig. \ref{Fig8}(d), the screw uniformly degrades as it is not subjected to mechanical loading. The dissolution process results in a reduction of both the outer and inner diameters of the shaft. Notably, increased volume loss is observed in the regions of the threaded screw and outer shaft. The upper part of the screw becomes disconnected from the threaded shaft region after forty days of immersion, Fig. \ref{Fig8}(d). This moment can be treated as the point at which the screw is no longer structurally responsible for load transfer. Complete dissolution of the screw without implant residue occurs after ninety-four days of immersion. It is emphasized that the degradation kinetics of the screws would be more pronounced in load-bearing implant systems as they are responsible for transferring loads between bone segments, Fig. \ref{Fig7}(a). The returned timescales in months for the dissolution of Mg plates and screws are consistent with existing literature \cite{LI2017, CHAYA2015, ZHAO2017}.

\subsection{Bioabsorbable Mg alloy porous scaffolds for bone tissue engineering}  \label{sec52}

\begin{figure}[h!]
    \centering
    \includegraphics[width = 16 cm]{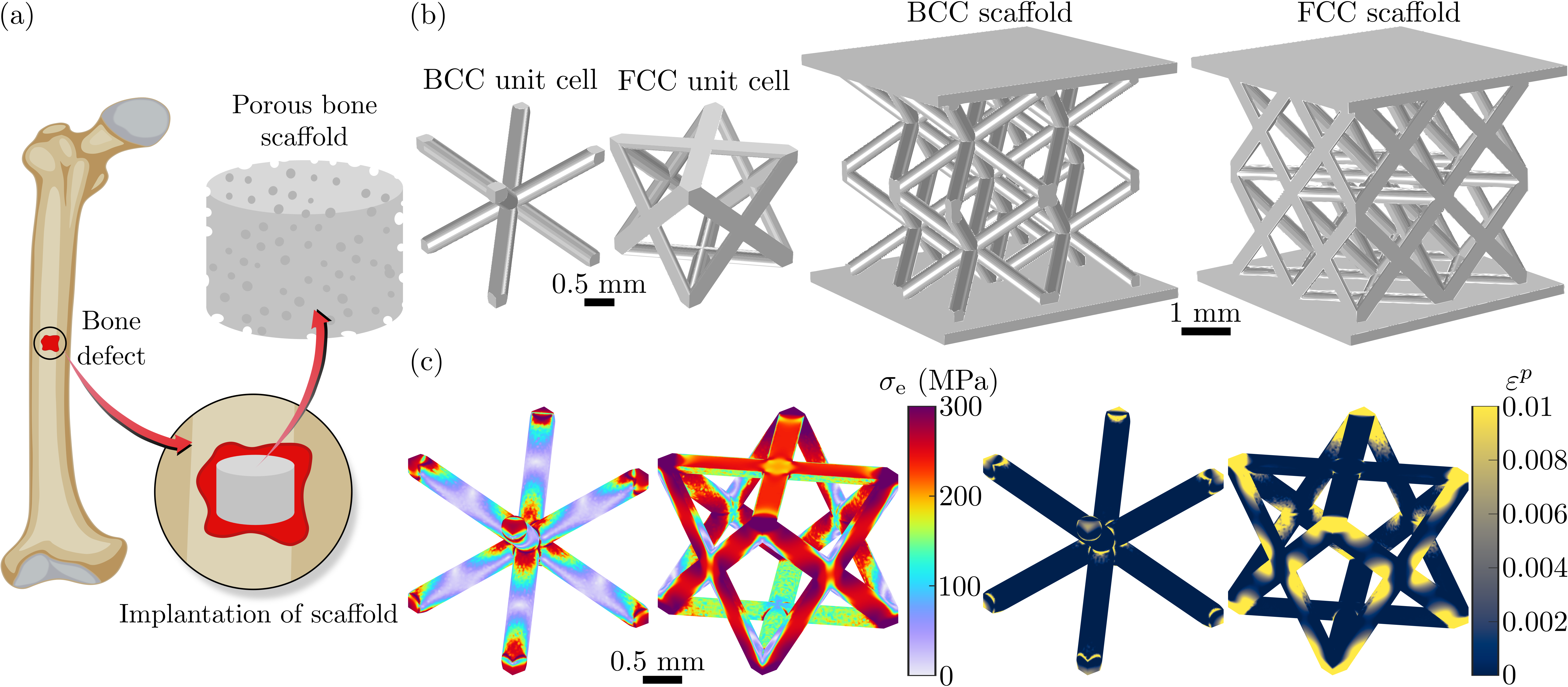}
    \captionsetup{labelfont = bf, justification = raggedright}
    \caption{Bioabsorbable Mg alloy porous bone scaffolds. (a) Representative example of porous bone scaffolds for critical-sized bone defects. (b) Body-centered cubic (BCC) and face-centered cubic (FCC) unit cell and scaffold structures considered in this work. (c) Distribution of von Mises stress $\sigma_{\mathrm{e}}$ and equivalent plastic strain $\varepsilon^p$ for the BCC and FCC scaffolds subjected to remote tensile strain $\varepsilon^\infty = 6$\%.}
    \label{Fig10}
\end{figure}

Bioabsorbable Mg alloy porous scaffolds are attractive for healing critical-sized bone defects that do not heal spontaneously within a patient's lifetime, Fig. \ref{Fig10}(a). These scaffolds act as bone substitutes and enable full regeneration of load-bearing bony defects \cite{LI2018b}. The problems mentioned above of rapid biodegradation kinetics of Mg alloys become even more crucial for porous scaffolds due to the enlarged free surface area by generating more complex geometries such as scaffold structures. The mechanical properties and corrosion resistance of porous scaffolds are typically determined in \textit{in vivo} and \textit{in vitro} tests \cite{KOPP2019, LIU2022}. It is demonstrated in this example that the present model can assess the performance of complex Mg implants, such as porous scaffolds, subjected to mechanical loading.  

This case study considers Mg alloy porous scaffolds with body-centered cubic (BCC) and face-centered cubic (FCC) unit lattice structures. Each scaffold consists of eight unit cells arranged in a 2x2x2 array. Each unit cell lattice has 2 mm in width and height. The lattice consists of cylindrical struts with a nominal strut diameter of 0.30 mm. Representative BCC and FCC unit cell and scaffold structures considered in this investigation are portrayed in Fig. \ref{Fig10}(b). The sizes of the unit cell and scaffold adopted in this study are commonly used in the literature \cite{LI2021}. The electrolyte size is chosen to be much larger than the size of the scaffolds to represent realistic implant conditions. It is assumed that the scaffolds are manufactured from the same Mg alloy as the one used in the \textit{in vitro} tests in Section \ref{sec2}. The mechanical properties of the alloy, the material, and the phase-field parameters follow those used in Section \ref{sec51}. Human blood conditions in terms of pH and initial concentration of Mg$^{2+}$ and Cl$^-$ ions are used to set up the initial conditions of ionic species. The pH is not controlled, following the decision made in the previous example. The boundary conditions for phase-field and concentration variables follow those used in the previous sections. 

Two different case studies are considered. In the first study, the BCC and FCC scaffolds are subjected to a constant remote tensile strain $\varepsilon^\infty = 6$\% applied in the vertical direction prior to immersion tests. The load magnitude is selected to induce plastic strains of around 1\% only in small areas within the structures. It is assumed that this level of plastic strain develops during the manufacturing and implantation processes. The distributions of von Mises stress and equivalent plastic strain after loading are displayed in Fig. \ref{Fig10}(c). The obtained plastic strains are then incorporated into the subsequent corrosion simulation. In the second study, the scaffolds are immersed in the corrosive medium in the absence of mechanical loading, representing idealized manufacturing and implantation conditions. The second case study represents uniform corrosion and serves as a reference for comparison.

\begin{figure}[h!]
    \centering
    \includegraphics[width = 16 cm]{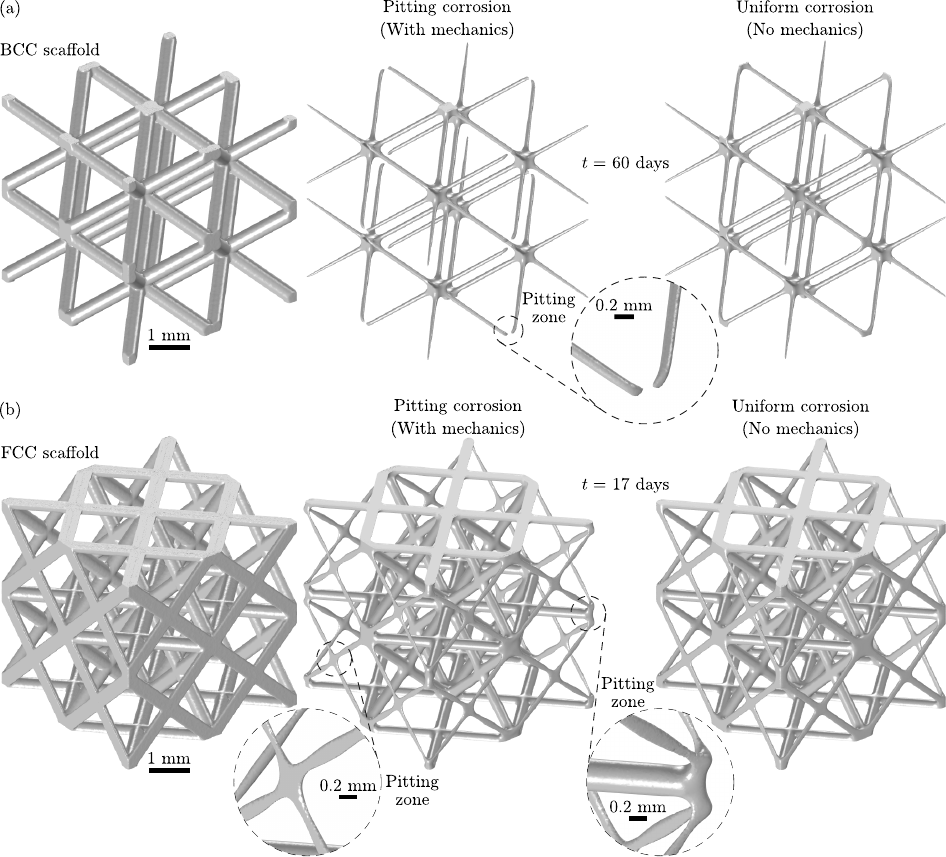}
    \captionsetup{labelfont = bf, justification = raggedright}
    \caption{Bioabsorbable Mg alloy porous bone scaffolds. Pitting and uniform corrosion of (a) BCC scaffolds after sixty days of immersion and (b) FCC scaffolds after seventeen days of immersion.}
    \label{Fig11}
\end{figure}

\begin{figure}[h!]
    \centering
    \includegraphics[width = 16 cm]{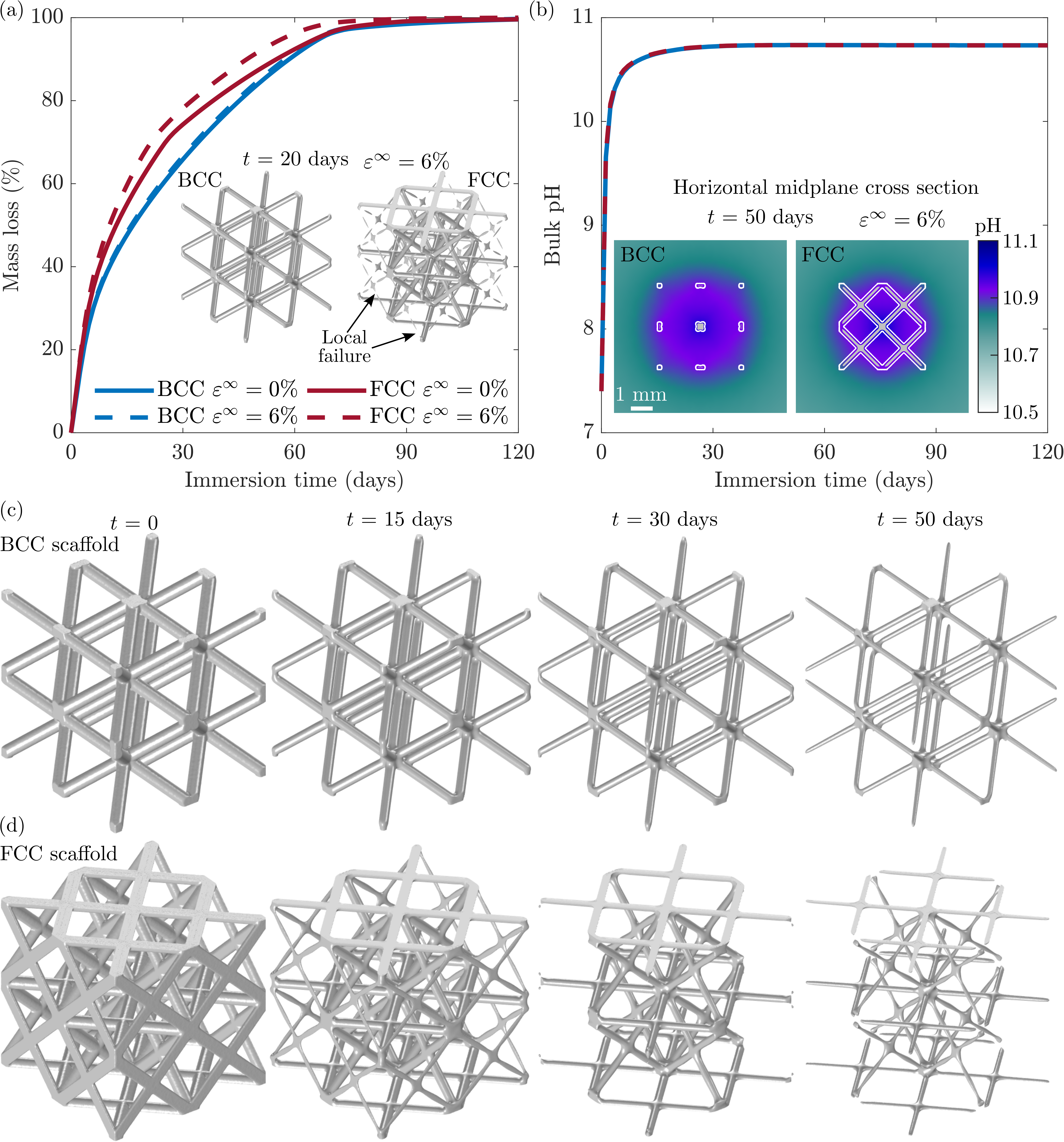}
    \captionsetup{labelfont = bf, justification = raggedright}
    \caption{Bioabsorbable Mg alloy porous bone scaffolds. (a) Mass loss and (b) bulk pH as a function of immersion time for the BCC and FCC scaffolds subjected to various remote tensile strain $\varepsilon^\infty$. Evolution of degradation of the (c) BCC and (d) FCC scaffolds subjected to the remote tensile strain $\varepsilon^\infty = 6$\% as a function of immersion time. The white line in the inset in (b) indicates the initial metal$-$electrolyte interface. The legend in the inset in (b) applies to the whole computational domain.}
    \label{Fig12}
\end{figure}

The obtained results in terms of pitting (considering mechanics) and uniform corrosion (without mechanics) for both BCC and FCC scaffolds are shown in Fig. \ref{Fig11}. Both scaffolds subjected to mechanical loading experience localized corrosion immediately upon immersion due to the initial plastic strains. However, different pitting zones are observed in the BCC and FCC scaffolds. A pitting zone is noted at the corners of each unit cell in the BCC scaffolds. FCC scaffolds show pitting zones at both the corners of each unit cell and the center of each face. The locations of these pitting zones correspond to the position of initial values of plastic strains. The local dissolution rate within these pitting zones is much higher than in the remaining areas of the scaffold. Although this highly localized corrosion has a negligible effect on overall mass loss and pH (Fig. \ref{Fig12}(a) and Fig. \ref{Fig12}(b)), it significantly impacts the mechanical integrity and load-bearing capacity of the scaffolds. The local deterioration of the thickness of the strut leads to the formation of defects within the structures. Consequently, connectivity between the unit cells disappears and large voids form with prolonged immersion time. Thus, the locations of the pitting zones indicate potential hot spots for early scaffold failures. The service life of the BCC scaffolds under mechanical loading is compromised after sixty days of immersion, while the integrity of the FCC scaffold is undermined after just seventeen days of immersion. The evolution of degradation of the BCC and FCC scaffolds subjected to the remote tensile strain $\varepsilon^\infty = 6$\% as a function of immersion time is depicted in Fig. \ref{Fig12}(c) and Fig. \ref{Fig12}(d). On the other hand, uniform corrosion progresses more gradually and affects the entire scaffold with a constant dissolution rate. The predicted service life for both BCC and FCC scaffolds in the case of uniform corrosion is longer than that for pitting corrosion. The simulations show that the BCC scaffold exhibits better corrosion resistance compared to the FCC scaffold. 

\section{Discussion} \label{sec6}

The current \textit{in vitro} study demonstrates that variations in chloride content, within a range relevant to biomedical applications, do not affect the corrosion performance of WE43MEO Mg alloys. This finding suggests that the well-accepted tendency that an increase in chloride concentration leads to higher corrosion rates \cite{Atrens2015_review, ZHAO2008, Wolf_Dieter2009, Johnston_Atrens2018_review, XIN2008_agressiveIons, WANG2010_chloride} does not apply to chloride concentrations typically used in corrosive media for testing Mg alloys for biomedical use \cite{MEI2020}. The present work highlights that pH plays a dominant role in Mg corrosion as it impairs the integrity and compactness of the corrosion product layer, leading to accelerated corrosion rates. Significantly higher corrosion rates are observed in a buffered solution compared to the air (buffer-free) environment. Hence, the likely cause of different corrosion rates among various corrosive media \cite{MUELLER2010, Xin_2010} may be primarily attributed to differences in pH and concentrations of inorganic ions, organic compounds, and proteins.

The observed pH dominance and electrochemical mechanisms by which pH affects corrosion rates can be possibly attributed to three combined factors: the formation of the magnesium hydroxide layer, shifts in the equilibrium electrode-electrolyte potential difference, and the rate of hydrogen evolution reaction. The effectiveness of the hydroxide layer that forms on the exposed metal surface is highly dependent on the pH of the surrounding medium. Buffer-free solutions with a pH around 10.5 tend to form a stable and highly compact Mg(OH)$_2$ layer, which slows the ingress of the solution toward the corroding metal surface, subsequently impeding its further degradation. Despite the bulk pH being around 6.2 in buffer-regulated media, the results indicate that the pH near the metal surface is higher than that of the remote solution, as shown in the insets in Fig. {\ref{Fig5}}(d). Even though the Mg(OH)$_2$ layer is not thermodynamically stable at low pH values (Fig. {\ref{Fig3}}(b)), the increased interfacial pH observed in buffer-controlled solutions stimulates the precipitation of the hydroxide layer, as illustrated in Fig. {\ref{Fig6}}(c). However, this interfacial pH remains lower than in buffer-free environments, leading to a less stable Mg(OH)$_2$ layer with larger voids and imperfections. This loose layer structure results in lower density and a propensity for the layer to detach from the metal surface. This promotes faster dissolution of the layer and exposes bare metal to a corrosive environment. The implication is that the fraction of layer-free surface increases with decreasing interfacial pH, raising the likelihood of higher corrosion rates and the occurrence of localized (pitting) corrosion. The change in interfacial pH between buffer-free and buffer-controlled solutions also affects the equilibrium electrode-electrolyte potential difference, Eq. ({\ref{eqn15}}). At low pH, the electrochemical potential becomes more negative, accelerating the corrosion process and returning higher corrosion rates. The potential difference is less negative at high pH values, yielding reduced corrosion rates. In addition, pH influences the rate of hydrogen evolution reaction. In solutions with low pH, hydrogen ions are abundant and can readily accept electrons from the metal surface to form hydrogen gas, see Eq. ({\ref{eqn1b}}). This reaction is particularly rapid in acidic environments, further enhancing the corrosion rate. In contrast, in solutions with high pH, hydrogen evolution is less pronounced due to the lower concentration of hydrogen ions. The individual contributions of these three factors to the electrochemical mechanisms by which pH affects corrosion rates are still not fully understood.

The present mechano-electrochemical-coupled phase-field framework for assessing localized corrosion and stress corrosion cracking in bioabsorbable Mg alloys integrates key physical phenomena associated with their corrosion in biological fluids. Distinguishing features of this model compared to existing similar approaches in the literature \cite{KOVACEVIC2023, XIE2024, ZHANG2024, OKAJIMA2010, Liang2012, Mai2018, Chadwick2018, Ansari2018, Tsuyuki2018, LIN2021, Brewick2022, Cui2023, Makuch2024, KANDEKAR2024} include its ability to account for (i) both anodic and cathodic reactions, (ii) the simultaneous formation and dissolution of the corrosion product layer, (iii) sensitivity to pH and chloride concentration, (iv) both buffer-free and buffer-regulated solutions, and (v) the role of mechanical fields in accelerating corrosion kinetics. The driving force for interface migration (Eq. (\ref{eqn12})) is decomposed into two contributions: the reduction of interfacial free energy and the electrochemical reaction at the metal$-$electrolyte interface, an approach followed in the literature \cite{Liang2012, Chadwick2018, Tsuyuki2018, LIN2021, Brewick2022}.  The first term on the right-hand side of Eq. (\ref{eqn12}) accounts for the influence of interfacial energy on the evolution of the phase-field parameter. The second term represents the electrode kinetics induced by corrosion. In the present framework, the evolution of the electrode kinetics is directly linked to the current density through Faraday’s law, Eq. (\ref{eqn13}). The current density is expressed using the standard Butler$–$Volmer equation (\ref{eqn14}), which includes three physical parameters: $i_0$, $\alpha_\mathrm{a}$, and $E^{\theta}$. All three parameters can be obtained experimentally using polarization curves. Alternatively, experimentally obtained current density curves can be utilized instead of the Butler$–$Volmer equation. The dependence of corrosion on pH is integrated into the present model by following the standard chemical thermodynamic relationship for the electrochemical reaction (Nernst equation), Eq. (\ref{eqn15}). The influence of mechanical fields on corrosion kinetics is incorporated into the model by following Gutman's mechanochemical theory {\cite{Gutman1988}}. This approach enhances the overpotential of the electrode, which in turn raises the current density, Eqs. ({\ref{eqn14}})$-$({\ref{eqn16}}). A higher current density is ascribed to increased corrosion rates. The final expression for the current density of the mechanically deformed electrode is provided in Eq. ({\ref{eqn16}}). This expression, along with Faraday's law (Eq. ({\ref{eqn13}})), is then used to describe the evolution of the solid$-$liquid interface, Eq. ({\ref{eqn17}}). The expression for current density can be simplified by neglecting the cathodic current density (due to the high standard electrode potential of Mg) and setting $\alpha_\mathrm{a} = 1$ \cite{KOVACEVIC2023}. The present model can be further simplified by excluding the concentration of Cl$^-$ ions as an independent kinematic variable. They do not affect the corrosion rate within the chloride range considered and their amount remains unchanged in the system ($R_3^{\prime} = R_3^{\prime\prime} = 0$) due to the assumption that MgCl$_2$ instantaneously dissolves in $\text{Mg}^{2+}$ and 2Cl$^-$, Eq. (\ref{eqn1d}).

The dimensional analysis (Section \ref{sec34}) indicates that the problem is governed by three non-dimensional parameters: $\tau_\phi, P_e$, and $D_a$. The selection of the first parameter $\tau_\phi \ll 1$ is based on the condition that the characteristic time for interface migration due to the reduction of interfacial energy is significantly larger than the characteristic times for the interface motion due to the electrochemical reaction. This condition provides the relationship for the phase-field mobility $L \ll v/ (\ell \omega)$, which implies that the interface movement is dominantly driven by the electrochemical reaction at the solid$-$liquid interface. The parameter $D_a$ controls the balance between the formation and dissolution of the Mg(OH)$_2$ layer. $P_e = \ell v/D_1^l$ determines the rate of the electrode reaction and activates different corrosion modes, including activation-controlled corrosion ($P_e \ll 1$) and diffusion-controlled corrosion ($P_e \gg 1$). $D_a$ is fitted to experimental measurements of the concentration of Mg ions in solution, Fig. \ref{Fig5}(c). The advantage of the current model lies in its reliance on fitting $D_a$ and physical parameters. Being able to obtain corrosion rates from physical parameters translates to fewer required model parameters, which consequently streamlines the modeling process while alleviating model calibration.

The methodology developed to capture the mechano-electrochemical effect is showcased through case studies focused on orthopedic applications. The simulations in Section \ref{sec51} illustrate that mechanical fields cause accelerated degradation rates in areas with high mechanical stresses and plastic strains. This locally enhanced degradation initiates crack formation and propagation, dramatically reducing the load-bearing capacity of the implant and leading to a loss of its structural integrity. A comparison between localized corrosion triggered by mechanical fields and uniform corrosion reveals that mechanics greatly influences the service life of implants with a negligible impact on mass loss and pH. Hence, relying solely on measuring mass loss and pH can provide misleading data on implant performance, overestimating the service life of the implant and concealing accelerated failure of load-bearing implants. The second case study in Section \ref{sec52} investigates the effect of unavoidable residual strains that occur after implantation and manufacturing processes on the performance of porous scaffolds. The results indicate that degradation starts with localized corrosion due to initial plastic strains and intensifies in high-strain regions, promoting pitting corrosion. Similar to the first case study, the presence of plastic strains shortens the service life of porous bone scaffolds. These examples in Section \ref{sec5} emphasize the importance of including mechanical fields in the design of orthopedic implants. The current investigation suggests that an ideal design that reduces the formation of stress concentrations can prolong the service life of orthopedic implants. Stress concentrations typically occur in areas with abrupt changes in geometry, resulting in localized regions of high stresses. Over time, these high stresses can lead to pit formation, pit-to-crack initiation, and crack propagation, ultimately causing failure under mechanical loading, as shown in the case studies in Section {\ref{sec5}}. By designing implants that distribute mechanical loads more uniformly, the risk of these stress concentrations can be minimized. Moreover, adjusting the shape of the implant to align with the specific mechanical needs of the bone tissue being replaced can further enhance its load-bearing capacity. This application-tailored implant design would ensure that the implant does not endure excessive mechanical loading, which could result in premature failure. 

While the model is not without its limitations, its sensitivity to pH, chloride concentration, and mechanical fields offers a robust basis for predicting the performance of bioabsorbable implants under realistic physiological conditions. Its effectiveness in simulating timescales comparable to actual implant lifespans makes it an invaluable tool for assessing implant behavior. With these advantages, the framework has the capability to influence the design of orthopedic implants. It provides a cost-effective approach for developing Mg-based implants and porous scaffolds with optimized topology and pore structure that achieve degradation rates synchronized with bone/tissue growth. However, the computational cost associated with 3D simulations remains a primary limitation. Certain assumptions are made in the present study to alleviate the computational demands to showcase the potential of the model. Future work should focus on developing advanced numerical algorithms for efficient and parallelized computations to address these challenges. The model can be expanded by incorporating additional chemical reactions and ionic species present in complex corrosive media. Extending the chemistry framework and integrating environmental components would provide deeper insights into the effect of solution composition on the corrosion of Mg alloys. In addition to environmental factors, future work should consider the role of microstructural features, such as grains, grain boundaries, secondary phases, and impurities in the Mg matrix in the corrosion of these alloys.

\section{Conclusions} \label{sec7}

The degradation behavior of WE43MEO Mg wires is systematically investigated in air and buffer-controlled test environments with varying chloride concentrations. Compared to air conditions, the study shows that buffer-regulated solutions with a low pH return higher degradation rates. The investigation consistently demonstrates that variations in chloride concentration, relevant to biomedical applications, have a negligible effect on corrosion kinetics in both test environments. The findings reveal that pH is a dominant factor and plays a significant role in the corrosion of bioabsorbable Mg alloys.

A computational framework based on the phase-field method is constructed for assessing the degradation of bioabsorbable Mg implants in biological fluids. The model is applicable to both buffer-free and buffer-regulated media. The dependence of corrosion on pH, chloride concentration, and mechanical fields is integrated into the model. The framework is validated against the \textit{in vitro} tests on Mg wires in fluids with different pH and chlorine ion concentrations. Good agreement between experiments and simulations is attained. The importance of including mechanical fields in the design of bioabsorbable implants is highlighted in representative case studies considering Mg plates and screws for bone fracture fixation and porous scaffolds for healing critical-sized bone defects. The simulations indicate that mechanical loading initiates crack formation and propagation, reducing the load-bearing capacity and shortening the service life of Mg plates. Ignoring mechanical loading on implant dissolution provides misleading data on implant performance, overestimating the service life of implants and concealing accelerated failure of load-bearing implants. The study found that local plastic strain developed during implantation and manufacturing bone scaffolds act as initiators for pitting corrosion and hot spots for early implant failures. The results reveal that BCC scaffolds exhibit better corrosion resistance than FCC scaffolds. The constructed model can be a cost-effective tool for optimizing the performance and predicting the lifespan of temporary Mg-based biomedical devices.

%\vspace{-5mm}
\section*{Acknowledgments} \label{sec8}

S.K. and E.M.-P. acknowledge financial support from UKRI’s Future Leaders Fellowship program [Grant MR/V024124/1]. W.A. and J.LL. acknowledge financial support from the BIOMET4D project (Smart 4D biodegradable metallic shape-shifting implants for dynamic tissue restoration) under the European Innovation Council Pathfinder Open call, Horizon Europe Research and Innovation program, grant agreement No. 101047008, and from the Spanish Research Agency through the grant PID2021-124389OB-C21. T.K.M. acknowledges the Newton International Fellowship (NIF/R1/221159) funded by The Royal Society. The authors would like to acknowledge the use of the University of Oxford Advanced Research Computing (ARC) facility in carrying out this work (\url{http://dx.doi.org/10.5281/zenodo.22558}).
%\vspace{-5mm}
\section*{Data availability} \label{sec9}

Supplementary research data associated with this article can be found in the online version. The experimental data from Section 2, code input files for the case studies in Section 4, and the original geometry of the bioabsorbable Mg screw (Section 5.1) and BCC and FCC bone scaffolds (Section 5.2) are provided. The code developed together with example case studies and documentation will be available at \url{https://mechmat.web.ox.ac.uk/codes} after article acceptance.

\end{document}